\documentclass[aps,prb,twocolumn,groupedaddress]{revtex4}
\usepackage{graphicx}

\begin{document}


\title{Fourier analyses of commensurability oscillations in Fibonacci lateral superlattices}


\author{Akira Endo}
\email[]{akrendo@issp.u-tokyo.ac.jp}
\homepage[]{http://iye.issp.u-tokyo.ac.jp/endo/}
\author{Yasuhiro Iye}
\affiliation{Institute for Solid State Physics, University of Tokyo, 5-1-5 Kashiwanoha, Kashiwa, Chiba 277-8581, Japan}

\date{\today}

\begin{abstract}
Magnetotransport measurements have been performed on Fibonacci lateral superlattices (FLSLs) --- two-dimensional electron gases subjected to a weak potential modulation arranged in the Fibonacci sequence, $LSLLSLS...$, with $L/S$=$\tau$ (the golden ratio). Complicated commensurability oscillation (CO) is observed, which can be accounted for as a superposition of a series of COs each arising from a sinusoidal modulation representing the characteristic length scale of one of the self-similar generations in the Fibonacci sequence. Individual CO components can be separated out from the magnetoresistance trace by  performing a numerical Fourier band-pass filter. From the analysis of the amplitude of a single-component CO thus extracted, the magnitude of the corresponding Fourier component in the potential modulation can be evaluated. By examining all the Fourier contents observed in the magnetoresistance trace, the profile of the modulated potential seen by the electrons can be reconstructed with some remaining ambiguity about the interrelation of the phase between different components.
\end{abstract}


\maketitle




%

\section{Introduction}
A GaAs/AlGaAs-based high mobility two-dimensional electron gas (2DEG) combined with modern electron-beam (EB) lithography allows us to fabricate a device in which electrons travel through an artificially designed environment over a sufficiently long distance before being scattered by impurities. \cite{BeenakkerR91} A prototypical example of such devices is a lateral superlattice (LSL), where a periodic modulation of electrostatic potential is introduced to a 2DEG\@. The periodic modulation can be either one-dimensional (1D) or two-dimensional (2D), exemplifying the versatility of LSLs compared with more conventional vertical superlattices. The planar geometry of a LSL is well-suited for experimental studies of the transport properties, and therefore the magnetotransport of both 1D and 2D LSLs, with the magnetic field $B$ applied perpendicular to the 2DEG plane, have been the subject of extensive studies [see, e.g., Refs. \onlinecite{Weiss89,Winkler89,Beton90P,Geim92,Muller95,Tornow96,Milton00,Endo02f,Endo04EP,Smet99c,Willett99c,Endo01c,Deutschmann01,Endo05N} and Refs. \onlinecite{Gerhardts91,Schlosser96,Albrecht99,Geisler04,Chowdhury04} (and references therein) for 1D and 2D LSLs, respectively].

In a 1D-LSL, a number of phenomena are known in the magnetotransport, including low-field positive magnetoresistance (PMR), \cite{Beton90P} commensurability oscillation (CO) originating from the commensurability between the cyclotron radius and the modulation period, \cite{Weiss89} quantum interference of closed orbits, \cite{Deutschmann01} and geometric resonance of open orbits. \cite{Endo05N} The periodic modulation in a 1D-LSL can be replaced by a quasiperiodic one without appreciable technical difficulties. Quasiperiodic structures are characterized by long-range order without translational symmetry, and constitute intriguing intermediate states situated between perfect periodicity and disorder. \cite{NoteRev} A wealth of exotic properties, such as localized or critical wave functions and a Cantor-set spectrum, are theoretically predicted, \cite{Aubry80,Sokoloff85,Kohmoto87} with varieties of experimental systems devoted for their exploration. \cite{Merlin85,Todd86,Yamaguchi90,Gellermann94,Munzar94,Toet91,Mikulik95,Zhu89,Macon91,Kono91,Katsumoto93,Smith96,Ebert99,Moras06,Eom06} It is therefore of great interest to perceive how the magnetotransport phenomena observed in a periodic 1D-LSL are altered (or remain unaltered) in a quasiperiodic 1D-LSL, as well as to search for phenomena peculiar to the quasiperiodic systems. The Fibonacci sequence typifies quasiperiodic systems in one-dimension. The present authors have been studying the magnetotransport of LSL samples with the potential modulation arranged in the Fibonacci sequence --- the Fibonacci LSL (FLSL). In previous publications, we briefly reported complicated CO and geometric resonance of open orbits observed in FLSLs, \cite{Endo07I} as well as the enhancement of zero-field resistivity followed by strong negative magnetoresistance for FLSLs with short-length-scale modulation approaching the Fermi wave length. \cite{Endo08E} In the present paper, we focus on the quantitative Fourier analysis of CO. We will show that the magnitudes of individual Fourier components that constitute the potential modulation can be determined by the analysis. Detailed knowledge of the potential profile thus obtained forms the basis for understanding the electronic properties or phenomena in FLSLs to be explored in the future studies. 

The paper is organized as follows. In Sec.\ \ref{Periodic}, the CO for a simple sinusoidal periodic LSL and its extension to systems with multiple Fourier components are briefly reviewed, in order to prepare for later application to the analysis of FLSLs. Details of the samples and the experimental setup are described in Sec.\ \ref{Experiment}. The main subject of the paper is presented in Sec.\ \ref{FA}, where we delineate the prescription to single out CO corresponding to a particular Fourier component from the magenetoresistance trace of a FLSL, and to deduce the amplitude of the potential modulation for the component; we further make an attempt to reconstruct the potential profile by exploiting the phases of the Fourier components inferred by examining the origin of each component. Discussion on the limitation in the accuracy of the potential profile thus determined is given in Sec.\ \ref{Discussion}, followed by concluding remarks in Sec.\ \ref{Conclusions}.

\section{Commensurability oscillation in periodic lateral superlattices \label{Periodic}}
In a periodic 1D-LSL, the oscillatory component of the resistivity in the CO, $\delta \rho_{xx}^g(B)$, is well understood quantitatively. We take the $x$-direction to be the direction of the modulation $V(x)$ and also the direction of the current for the resistivity measurement, and the $xy$-plane to be the plane of the 2DEG with $B$ applied along the $z$-axis, throughout this paper. For a sinusoidal modulation $V(x)=V_g \cos(g x)$ with period $a$=$2\pi/g$ and with the amplitude $V_g$ much smaller than the Fermi energy $E_\mathrm{F}$, $\delta \rho_{xx}^g(B)$ has been experimentally shown to be accurately described by a simple analytic formula, \cite{Endo00E,NoteBsign}
\begin{equation}
\frac{\delta \rho_{xx}^g(B)}{\rho_0 A(T/T_g)}=\gamma A\left( \frac{\pi}{\mu_\mathrm{w}B} \right) g V_g^2 B \sin (2 g R_\mathrm{c}),
\label{magres1}
\end{equation}
with
\begin{equation}
\gamma=\frac{1}{2(2\pi)^{3/2}}\left(\frac{h}{e}\right)^{-1}\left(\frac{e\hbar}{2m^*}\right)^{-2}\frac{\mu^2}{n_\mathrm{e} ^{3/2}},
\label{cnst}
\end{equation}
$\rho_0$=$\rho_{xx}(0)$, $A(X)$=$X/\sinh(X)$, $k_\mathrm{B}T_g$=$(1/2 \pi)(k_\mathrm{F}/g)\hbar\omega_\mathrm{c}$, $m^*$ the electron effective mass (0.067$m_e$ for GaAs), $\omega_\mathrm{c}$=$eB/m^*$ the cyclotron angular frequency, $R_\mathrm{c}$=$\hbar k_\mathrm{F}/eB$ the cyclotron radius, $k_\mathrm{F}$=$\sqrt{2\pi n_e}$ the Fermi wave number, and $\mu$ and $n_e$ represent the mobility and the density of the electrons, respectively. The factor $A(\pi/\mu_\mathrm{w}B)$ in Eq.\ (\ref{magres1}) is the damping factor arising from the scattering of electrons out of their cyclotron orbit, with $\mu_\mathrm{w}$ an appropriate mobility usually identifiable with the quantum mobility $\mu_\mathrm{q}$ that describes the damping of the Shubnikov-de Haas (SdH) oscillation. Apart from the thermal and scattering damping factors $A(T/T_a)$ and $A(\pi/\mu_\mathrm{w}B)$, Eq.\ (\ref{magres1}) can be deduced from a physically transparent semiclassical picture. \cite{Beenakker89} Electrons acquire a drift velocity $v_{\mathrm{d},y}(x)$=$-E(x)/B$ in the simultaneous presence of an electric field $E(x)$=$(1/e)dV(x)/dx$ and a magnetic field $B$. The average of the drift velocity over one cycle of the cyclotron orbit
\begin{eqnarray}
\overline{v_{\mathrm{d},y}}(x_0) = \frac{1}{2\pi}\int_0^{2\pi } \!\!\! d\theta {v_{\mathrm{d},y} } (x_0+R_\mathrm{c} \cos \theta ) \nonumber\\
=\frac{g V_g}{eB} \sin (g x_0) J_0 (g R_\mathrm{c})
\end{eqnarray}
oscillates with $B$ reflecting the commensurability between $a$ and $R_\mathrm{c}$, with the amplitude depending on the guiding center position $x_0$. Here, $J_0(X)$ represents the Bessel function of order 0. The conductivity deriving from the commensurability effect, $\delta \sigma_{yy}$, is obtained by further averaging the square $\overline{v_{\mathrm{d},y}}^2(x_0)$ over $x_0$ (averaging over one modulation period suffices for a periodic LSL), and substituting the resultant $\langle \overline{v_{\mathrm{d},y}}^2 \rangle$ in the Einstein's relation, $\delta \sigma_{yy}$=$e^2(m^*/\pi \hbar^2)( \tau_\mathrm{m} \langle \overline{v_{\mathrm{d},y}}^2 \rangle)$, with $\tau_\mathrm{m}$=$(e/m^*)\mu$ the (momentum relaxation) scattering time. The corresponding resistivity is obtained by inverting the conductivity tensor: $\delta \rho_{xx}/\rho_0$=$(\omega_\mathrm{c}\tau_\mathrm{m})^2\delta \sigma_{yy}/\sigma_0$ with $\sigma_0=n_ee^2\tau_\mathrm{m}/m^*$ for $(\omega_\mathrm{c}\tau_\mathrm{m})^2$=$(\mu B)^2$$\gg$1. This, with the Bessel function replaced by its asymptotic expression $J_0(X)$$\simeq$$\sqrt{2/\pi X}\cos(X-\pi/4)$,  coincides with Eq.\ (\ref{magres1}) except for non-oscillatory term linear in $B$ and the damping factors. The scattering damping factor can be incorporated within the semiclassical picture by the Boltzmann equation approach. \cite{Mirlin98} The thermal damping factor results from standard treatment of the blurring of the edge of the Fermi-Dirac distribution function. Virtually the same result (with $A(T/T_a)$ but without $A(\pi/\mu_\mathrm{w}B)$) is arrived at by quantum mechanical calculations that treat $V(x)$ by the first-order perturbation theory. \cite{Zhang90,Peeters92}

By fitting Eq.\ (\ref{magres1}) to an experimentally obtained CO trace, unknown parameters of the LSL sample under investigation can be deduced. It can readily be seen from Eq.\ (\ref{magres1}) that the oscillation is periodic in $1/B$ with minima taking place at the conditions,
\begin{equation}
\frac{gR_\mathrm{c}}{\pi}=\frac{B_g}{B}=n-\frac{1}{4}\hspace{10mm}(n=1,2,3,...).
\label{flatband}
\end{equation}
From the frequency $B_g$, the wave number $g$=$\pi eB_g/\hbar k_\mathrm{F}$ of the potential modulation (if unknown) is revealed using the $n_e$ determined either by the Hall resistivity or by the SdH oscillation. The parameters $V_g$ and $\mu_\mathrm{w}$ are obtained from the oscillation amplitude. Fitting the experimental values of $|\delta \rho_{xx}^g(B)/\rho_0 A(T/T_g)|$ at the extrema to the function $C/\sinh(B_\mathrm{w}/B)$ employing $C$ and $B_\mathrm{w}$ as fitting parameters, we get $V_g$=$\sqrt{C/(B_\mathrm{w} \gamma g)}$ and $\mu_\mathrm{w}$=$\pi/B_\mathrm{w}$, where $\gamma$ is evaluated by Eq.\ (\ref{cnst}) using experimentally obtained parameters $\mu$ and $n_e$.

Extension of the above argument to general potentials of the form $V(x)$=$\sum_g V_g \cos (gx+\phi_g)$ is straightforward. The resulting conductivity was shown, using the quantum mechanical Kubo formula, to be simply additive; \cite{Gerhardts92} it can also readily be shown by the semiclassical derivation presented above by noting that the cross-terms in the square of the cycle-averaged drift velocity $\overline{v_{\mathrm{d},y}}(x_0)$=$\sum_g (g V_g/eB) \sin (g x_0+\phi_g)J_0(g R_\mathrm{c})$ vanish upon averaging over $x_0$. The total oscillatory part of the CO thus reads
\begin{equation}
\delta \rho_{xx}^\mathrm{tot}=\sum_g \delta \rho_{xx}^g
\label{rhoadd}
\end{equation}
with $\delta \rho_{xx}^g$ given by Eq.\ (\ref{magres1}). The phase $\phi_g$ is equivalent to the shift of the guiding center by $\phi_g/g$, and therefore does not appear in Eq.\ (\ref{rhoadd}) obtained by the averaging over $x_0$; the information on the phase is lost in the resistivity. For a (non-sinusoidal) periodic potential that includes higher harmonics, \cite{Endo05HH} the $g$ in the summation contains only integer multiples of the fundamental component $g_0$, i.e., $g$=$\lambda g_0$ ($\lambda$=1,2,3,...). However, Eq.\ (\ref{rhoadd}) is applicable to more general potential that involves $g$'s with the ratio $\lambda$ given by irrational numbers, as is the case in our present FLSL samples. Once experimentally measured $\delta \rho_{xx}^\mathrm{tot}$ is decomposed into its constituent $\delta \rho_{xx}^g$'s, the corresponding $V_g$'s can be obtained following the prescription for a sinusoidal modulation described above. This is the main subject of the present paper that will be demonstrated in what follows.

\section{Experimental details \label{Experiment}}
\begin{figure}[tb]
\includegraphics[bbllx=15,bblly=65,bburx=810,bbury=370,width=8.5cm]{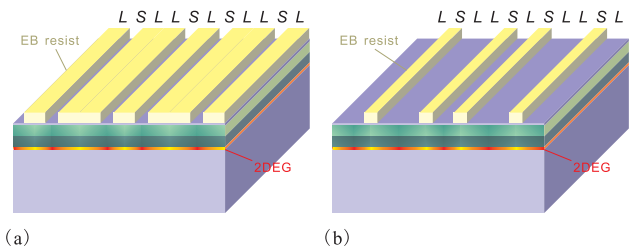}%
\caption{(Color online) Schematic illustrations of the FLSL samples. (a) and (b) represent $L$-type and $S$-type samples, respectively. \label{samplesch}}
\end{figure}

\begin{table}
\caption{List of samples. $\langle \Delta x_\mathrm{res} \rangle$ denotes the average distance between the centers of adjacent resist slabs. \label{Sampletbl}}
\begin{ruledtabular}
\begin{tabular}{ccccccc}
Sample ID & $L$ (nm) & $S$ (nm) & $\langle \Delta x_\mathrm{res} \rangle $ (nm) & Type \\
\hline
90$L$ & 104 & 64 & 231 & $L$ \\
70$L$ & 81 & 50 & 180 & $L$ \\
60$L$ & 69 & 43 & 154 & $L$ \\
55$L$ & 63 & 39 & 141 & $L$ \\
50$L$ & 58 & 36 & 129 & $L$ \\
45$L$ & 52 & 32 & 116 & $L$ \\
40$L$ & 46 & 28 & 103 & $L$ \\
110$S$ & 127 & 78 & 283 & $S$ \\
100$S$ & 115 & 71 & 257 & $S$ \\
90$S$ & 104 & 64 & 231 & $S$ \\
\end{tabular}
\end{ruledtabular}
\end{table}

The Fibonacci sequence, $LSLLSLSLLSL...$ , is generated by first preparing two unit lengths, $L$ and $S$ with the ratio $L/S$=$\tau$=$(1+\sqrt{5})/2$=1.61803... (the golden ratio), and then repeating the inflation rules, $S \rightarrow L$ and $L \rightarrow LS$, starting from a single $S$. In a FLSL, modulation arranged in the Fibonacci sequence is introduced to a 2DEG. The samples are fabricated from conventional GaAs/AlGaAs single heterostructure 2DEG wafers ($\mu$=50--100 m$^2$/Vs, $n_e$=1.7--3.0$\times$10$^{15}$ m$^{-2}$, varied by LED illumination) with the heterointerface residing at the depth of 90 nm from the surface. The electrostatic potential modulation is introduced by exploiting the strain-induced piezoelectric effect, \cite{Skuras97} the strain being generated upon cooling by the slabs of negative EB-resist \cite{Endo00E} placed on the surface selectively on either of the $L$-site ($L$-type FLSL) or $S$-site ($S$-type FLSL), as depicted in Fig.\ \ref{samplesch}. Seven $L$-type and three $S$-type samples with varying unit lengths, as tabulated in Table \ref{Sampletbl}, are examined. The accuracy of the actual width of the resist slabs is estimated from the scanning electron micrographs to be roughly within ten percent of the ideal values.

\begin{figure}[tb]
\includegraphics[bbllx=55,bblly=440,bburx=540,bbury=805,width=8.5cm]{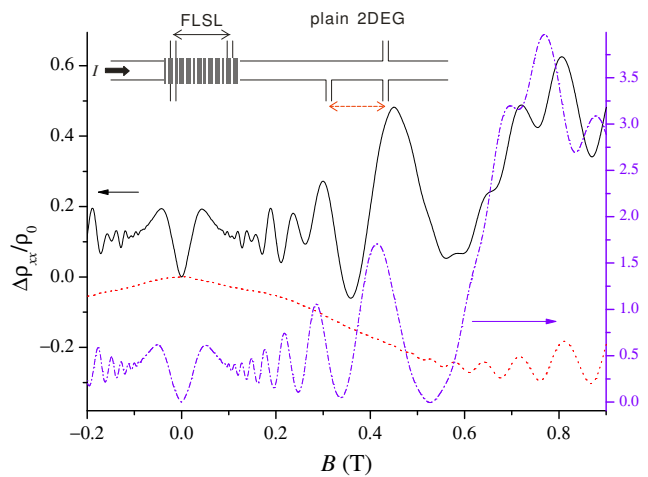}%
\caption{(Color online) Magnetoresistance traces of a FLSL (solid line) and the adjacent plain 2DEG (dotted line) taken at 4.2 K from sample 70$L$ (left axis). Magnetoresistance trace for a periodic LSL ($a$=184 nm, $V_g$=0.3 meV) is also shown for comparison (dash-dotted line, with negative offset, right axis). The inset schematically depicts the configuration of the sample. \label{FB28a4DGDFL}}
\end{figure}

A Hall-bar pattern with two sets of voltage probes are employed for the samples, as illustrated in the inset of Fig.\ \ref{FB28a4DGDFL}, allowing for the simultaneous measurement of the unpatterned plain 2DEG adjacent to the FLSL for reference. The width of the Hall bars and the center-to-center distance between the voltage probes are 14 $\mu$m and 40 $\mu$m, respectively. Thus the FLSLs contain from 140 (sample 110$S$) to 290 (sample 40$L$) segments of the $S$-site. The main panel of Fig.\ \ref{FB28a4DGDFL} shows a typical magnetoresistance trace of a FLSL, along with that of the adjacent plain 2DEG\@. The measurement is carried out by standard low-frequency ac lock-in technique. We have checked that the microfabrication process to introduce the modulation does not degrade the mobility and also does not alter the electron density (as is evident by comparing the SdH oscillations observed in the two traces for $B$$\agt$0.5 T). In the FLSL, the CO is observed for $B$$\agt$0.05 T, following the low-field PMR\@. The complicated non-monotonic nature of the CO in the FLSL is immediately evident by the comparison with the CO of a periodic LSL, also plotted in Fig.\ \ref{FB28a4DGDFL} for reference. Quantitative analysis of the CO in FLSLs will be presented in the following section.

\section{Fourier analyses of the commensurability oscillation \label{FA}}
\subsection{The separation of Fourier components and the analyses of the CO amplitudes}
 
\begin{figure}[tb]
\includegraphics[bbllx=50,bblly=80,bburx=520,bbury=800,width=8.5cm]{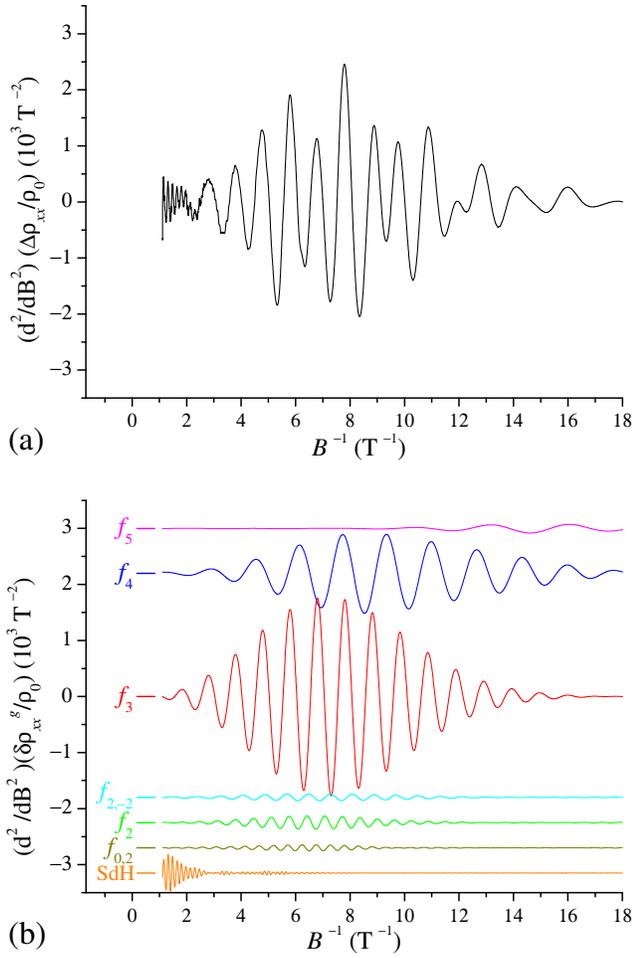}%
\caption{(Color online) (a) The plot of $(d^2/dB^2)(\Delta \rho_{xx}/\rho_0)$ vs. $1/B$. (b) Harmonic components of (a) obtained by performing Fourier band pass filters. Corresponding windows are shown in Fig.\ \ref{FB28a4FFT}. The left axis is for $f_3$, and other traces are offset for clarity. \label{FB28a4d2dBcns}} 
\end{figure}

\begin{figure}[tb]
\includegraphics[bbllx=60,bblly=440,bburx=520,bbury=830,width=8.5cm]{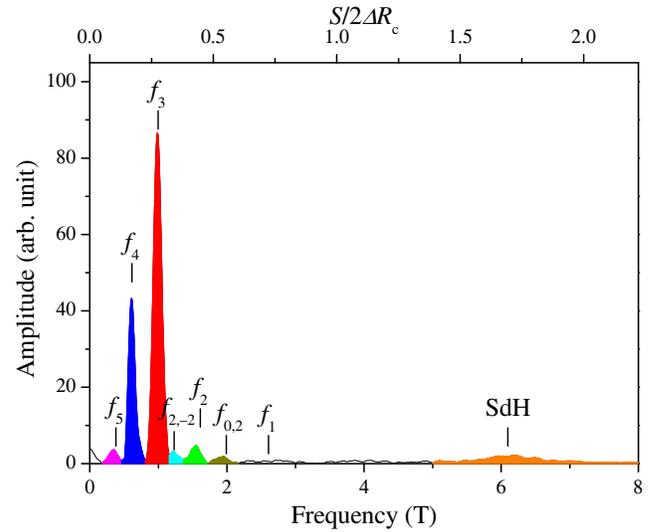}%
\caption{(Color online) The Fourier spectrum of the trace of $(d^2/dB^2)(\Delta \rho_{xx}/\rho_0)$ vs. $1/B$ shown in Fig.\ \ref{FB28a4d2dBcns} (a). The shaded areas indicate the windows employed in the band-pass filters in Fig.\ \ref{FB28a4d2dBcns}. \label{FB28a4FFT}}
\end{figure}

In this section, we describe, taking sample $70L$ as an example, the procedure to separate out Fourier components from a magnetoresistance trace $\Delta \rho_{xx}(B)/\rho_0$=$(\rho_{xx}(B)-\rho_0)/\rho_0$ as shown in Fig.\ \ref{FB28a4DGDFL}, and to analyze the CO amplitude of each Fourier content thus singled out. The first step is to subtract the slowly-varying background to obtain the oscillatory part. In the case of a periodic LSL, the background can readily be defined as the average of the upper and lower envelope curves; \cite{Endo00E} since the CO amplitude decreases monotonically with decreasing magnetic field, the envelope curves obtained as spline curves connecting maxima or minima, and hence their average, are also monotonic. A similar approach is not applicable for FLSLs having non-monotonic envelope curves. We, instead, make use of the numerical differentiation by $B$ that effectively operates as a high pass filter. The second derivative successfully eliminates the slowly-varying part, as can be seen in Fig.\ \ref{FB28a4d2dBcns}(a) which shows $(d^2/dB^2)[\Delta \rho_{xx}(B)/\rho_0]$ plotted against $1/B$. 

The Fourier transform of Fig.\ \ref{FB28a4d2dBcns}(a), presented in Fig.\ \ref{FB28a4FFT}, displays discrete peaks. This indicates that the beating of the oscillation in Fig.\ \ref{FB28a4d2dBcns}(a) can be viewed as resulting from the superposition of a number of components periodic in $1/B$. A peak originating from the SdH effect is also observed. The positions of the principal peaks marked in Fig.\ \ref{FB28a4FFT} by $f_j$ ($j$=5,4,3,2 and a small trace for $j$=1) are mutually related by $f_j/f_{j+1}=\tau$, namely, the positions of adjacent peaks are scaled by the golden ratio $\tau$. The other two minor peaks bear relations $f_{2,-2}=2f_4$ and $f_{0,2}=2f_3$. These peaks can be explained in terms of the reciprocal lattice for a 1D Fibonacci sequence. 

It is well known that quasiperiodic structures show a set of discrete sharp diffraction spots despite the absence of definite periodicity. \cite{Levine86} The reciprocal lattice of a 1D-Fibonacci lattice can readily be found by noting that the Fibonacci sequence can be generated also by the projection of a 2D-square lattice, with the lattice constant $\sqrt{1+\tau^2} S$, onto a line having the slope equal to $1/\tau$; the reciprocal lattice, or equivalently the position of peaks in the Fourier transform, is given by $g_{m,n}$=$(2\pi/S)f_{m,n}$ with
\begin{equation}
f_{m,n}=\frac{m\tau+n}{\tau+2}
\label{fmn}
\end{equation}
and $m$, $n$ integers. \cite{Elser86} Although in principle $g_{m,n}$'s fill up the entire 1D reciprocal space, only a small fraction of them with smaller values of $|m|$, $|n|$ have practically effective amplitudes of the structure factor, the detail of which depending on the actual profile of the potential. \cite{NoteStr}

The above-mentioned principal sequence coincides with Eq.\ (\ref{fmn}) with two successive Fibonacci numbers, $m$=$F_{3-j}$ and $n$=$F_{2-j}$,
\begin{equation}
f_j=f_{F_{3-j},F_{2-j}}=\frac{\tau^{3-j}}{\tau+2}=\frac{\tau^{2-j}}{\sqrt{5}},
\label{fj}
\end{equation}
if we translate the frequency $B_g$ into $S/2 \Delta R_\mathrm{c}$, the inverse of the corresponding increment in the cyclotron diameter $2 \Delta R_\mathrm{c}$=$2 \hbar k_\mathrm{F}/e B_g$ taking $S$ as the unit of the length (the top axis in Fig.\ \ref{FB28a4FFT}). Here the Fibonacci numbers $F_n$ are defined by $F_0$=0, $F_1$=1, and $F_{n+2}$=$F_{n+1}+F_n$ for both positive and negative integers $n$, and can be explicitly written as $F_n=[\tau^n-(-\tau)^{-n}]/\sqrt{5}$ (the Binet formula \cite{Dunlap97,Posamentier07}). In Eq.\ (\ref{fj}), we made use of an identity, $\tau^n$=$\tau F_n+F_{n-1}$. The two minor peaks are also expressed in terms of Eq.\ (\ref{fmn}).

The peak at $f_{m,n}$ corresponds to the modulation wave number $g$=$(2 \pi/S)f_{m,n}$ or the modulation period $S/f_{m,n}$. Therefore Fig.\ \ref{FB28a4FFT} reveals that the profile of the modulation is mainly composed of the superposition of multiple incommensurate components having the periods $S/f_j$, the periods successively scaled by an irrational number $\tau$. Physical interpretation of these periods will be presented in the following subsection.

\begin{figure}[tb]
\includegraphics[bbllx=45,bblly=100,bburx=520,bbury=800,width=8.5cm]{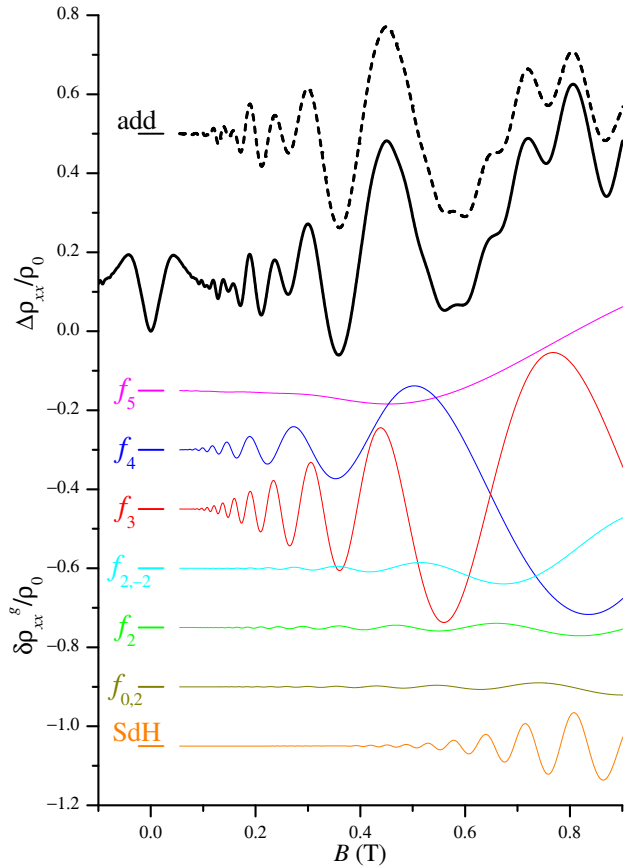}%
\caption{(Color online) Experimental magnetoresistance trace (thick solid line), the Fourier components extracted by the procedure described in the text (thin solid lines, with downward offsets), and the addition of the Fourier components (thick dashed line, with upward offset). \label{FB28a4cmps}}
\end{figure}

\begin{figure}[tb]
\includegraphics[bbllx=50,bblly=90,bburx=520,bbury=800,width=8.5cm]{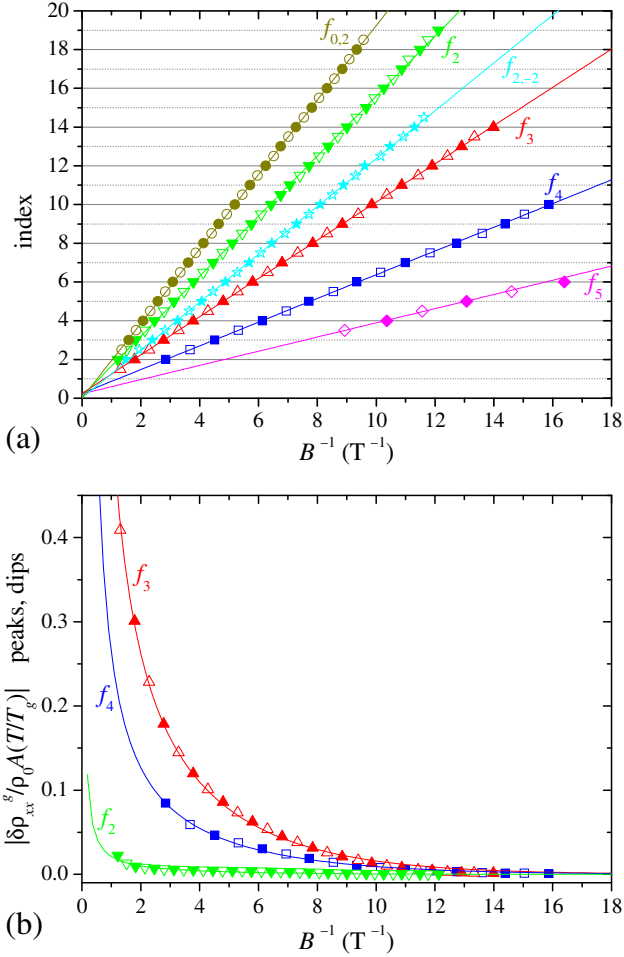}%
\caption{(Color online) (a) Plots of the index versus position (in the inverse magnetic field) of the extrema of CO Fourier components shown in Fig.\ \ref{FB28a4cmps}. The lines are linear fit to the data points. (b) Selected plots of the normalized amplitude versus position of the CO extrema. The curves are the fit to $C/\sinh(B_\mathrm{w}/B)$. Maxima and minima are plotted by open and solid symbols, respectively. \label{indampfit}}
\end{figure}

The trace in Fig.\ \ref{FB28a4d2dBcns} (a) can be decomposed into its Fourier components by performing a numerical Fourier band-pass filter. The resultant components obtained by applying the windows that encompass the corresponding peaks (shaded areas in Fig.\ \ref{FB28a4FFT}) are displayed in Fig.\ \ref{FB28a4d2dBcns} (b). The next step is to restore the resistivity $\delta \rho_{xx}^g$ from its second derivative. This can be readily done by numerically integrating the second derivative (the replot of Fig.\ \ref{FB28a4d2dBcns} (b) against $B$) by $B$ twice. The decomposed Fourier components of the CO (and also the SdH oscillation) thus obtained are plotted in Fig.\ \ref{FB28a4cmps}.  The summation of all the observed components (thick dashed curve) restores the oscillatory part of the original magnetoresistance trace (thick solid curve) with high fidelity, indicating that we are on the right track. As shown in Fig.\ \ref{indampfit} (a), the plot of the index of the extrema in each CO component versus their position in $1/B$ falls on a line with the slope $B_g$ that coincides with the peak position of the FFT spectrum (the lower axis in Fig.\ \ref{FB28a4FFT}) within experimental error, and with the ordinate intercept $\sim$1/4, as expected from Eq.\ (\ref{flatband}), thereby confirming that adequate CO components are extracted. Here we assigned half-integer indices to the maxima.

\begin{figure}[tb]
\includegraphics[bbllx=40,bblly=70,bburx=520,bbury=800,width=8.5cm]{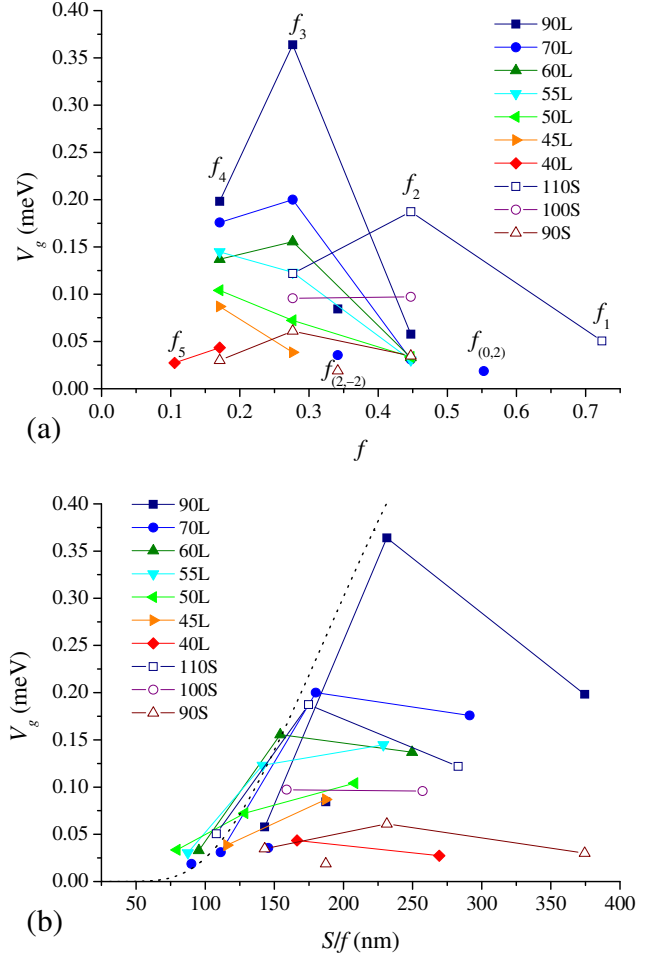}%
\caption{(Color online) (a) Amplitude of the potential modulation $V_g$ deduced from the CO amplitude for Fourier components specified by $f$. (b) Replot of (a) against the length $S/f$ corresponding to the Fourier component $f$. Dotted curve shows amplitude for periodic LSL, Eq.\ (\ref{expdecay1}). Solid and open symbols are used for $L$-type and $S$-type samples, respectively. Lines are guide for eyes and connect only the data points belonging to the principal sequence, $f_j$. \label{ampcns}}
\end{figure}

The final step is to deduce the modulation amplitude $V_g$ from the amplitude of CO following the procedure described in Sec.\ \ref{Periodic}. The values of $|\delta \rho_{xx}^g(B)/\rho_0 A(T/T_g)|$ at the extrema are plotted in Fig.\ \ref{indampfit} (b) for the three major components, $f_2$, $f_3$, and $f_4$. As can be seen, the function $C/\sinh (B_\mathrm{w}/B)$ describes the behavior of the normalized amplitudes quite well. Fitting by the two parameters $C$, $B_\mathrm{w}$ can be carried out without difficulty for the components with large enough amplitudes, such as $f_3$ and $f_4$ in the present case. The scattering parameter $\mu_\mathrm{w}$=$\pi/B_\mathrm{w}$ is expected not to vary among components. This is found to be the case for $f_3$ and $f_4$; the values obtained by the fitting, $\mu_\mathrm{w}$=10.2 and 11.3 m$^2$/Vs, respectively,  roughly coincide (within experimental error $\pm\sim $0.5 m$^2$/Vs). For components having smaller amplitudes as $f_2$ in Fig.\ \ref{ampcns} or other components not plotted in the figure, the two parameter fitting is found to be difficult to perform reliably. In such cases, we resort to the fitting by one parameter $C$ with a fixed $B_\mathrm{w}$ taken from the component with the largest amplitude ($f_3$ in the present case), relying on the invariability of the $\mu_\mathrm{w}$. Since only the ratio $C/B_\mathrm{w}$ is required for determining the $V_g$, possible small error in $B_\mathrm{w}$ does not affect the resulting $V_g$ very much, being compensated for by the responding change in $C$ obtained by the fitting. (Note that $C/\sinh(B_\mathrm{w}/B)$$\sim$$CB/B_\mathrm{w}$ for small $B_\mathrm{w}/B$.)

The values of deduced $V_g$ are plotted against $f$=$(S/2\pi)g$, namely, the values of $f_j$ or $f_{m,n}$ of each component given by Eqs. (\ref{fj}) and (\ref{fmn}), in Fig.\ \ref{ampcns} (a) for all the FLSL samples examined in the present work. As can be seen, dominant components are $f_4$, $f_3$, or $f_2$ in most samples. Minor components $f_{2,-2}$ and $f_{0,2}$ are occasionally detected for samples with longer unit lengths. An important factor that controls the modulation amplitude $V_g$ of the component $g$ is the corresponding length (the period) $2 \pi/g$=$S/f$ relative to the depth $d$ of the 2DEG plane, owing to the exponential decay $\propto \exp(-g d)$ along the depth of the effect exerted on the surface. In Fig.\ \ref{ampcns} (b), we replot $V_g$ against $S/f$. Dominant components for each sample are found to roughly follow the function that fits the modulation amplitude of periodic LSLs with period $a$=$2 \pi/g$ at $n_e$=2.8$\times$10$^{15}$ m$^{-2}$,
\begin{equation}
V_g=(-e)\phi_1 e^{-g d}/ {\epsilon_\mathrm{TF}}^\mathrm{FH}
\label{expdecay1}
\end{equation}
with $\phi_1$=43.8 meV, $d$=100 nm (the effective depth of the 2DEG that takes account of the average distance of the 2DEG wave function from the heterointerface), and ${\epsilon_\mathrm{TF}}^\mathrm{FH}$ representing the dielectric function of 2DEG in the Thomas-Fermi approximation using the Fang-Howard wave function. \cite{Endo05HH}

Now that we have determined $V_g$, the profile of the modulated potential can be reconstructed with further supply of the knowledge of the phases, $\phi_g$. Unfortunately, however, the information of the phases does not manifest itself in the CO, as mentioned earlier. We can only surmise the phase of Fourier components by looking into their origin, which is the subject of the subsequent subsection. Note however that in principle any phase relation between an arbitrary pair of components with irrational wave-number ratio can be found at some place in the sequence, provided that the sequence is infinitely long.

\subsection{The phase of the Fourier components}

\begin{figure*}[tb]
\includegraphics[bbllx=40,bblly=60,bburx=800,bbury=430,width=17cm]{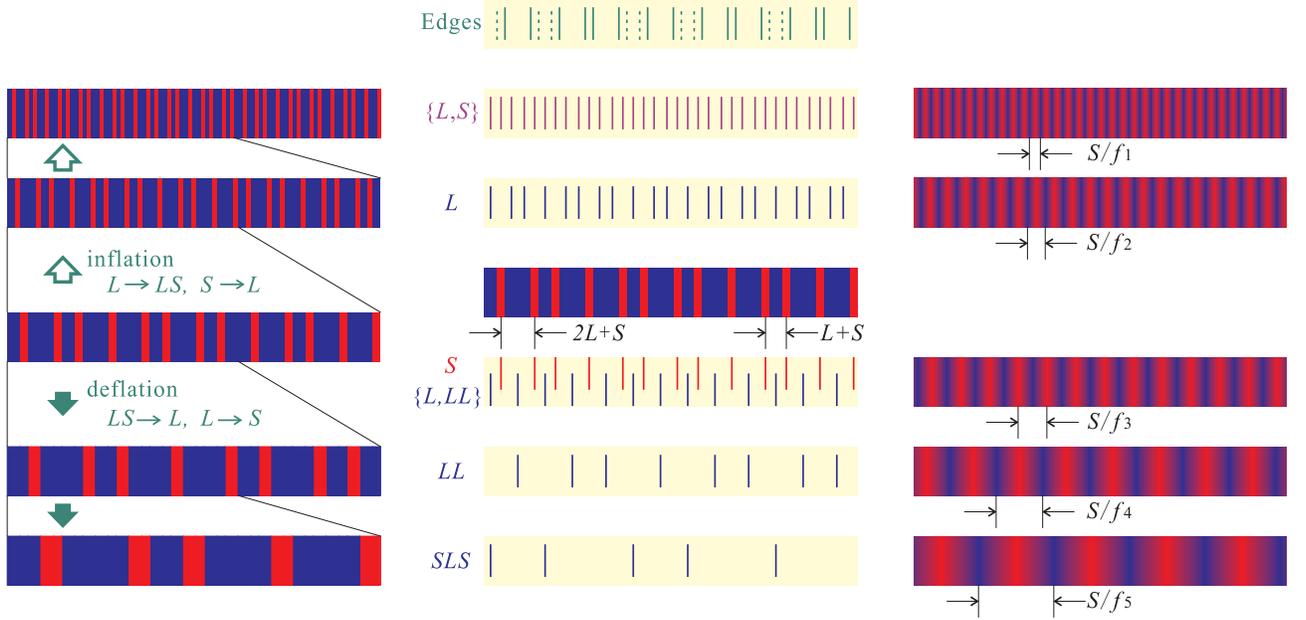}%
\caption{(Color online) Center: A diagram that sketches the Fibonacci sequence ($L$ and $S$ are presented by dark (blue) and bright (red) rectangles, respectively) with the vertical lines above and below indicating the locations of characteristic features: the centers of the segments not discriminating their types (\{$L$,$S$\}), the centers of the $L$ segments ($L$), the centers of the $S$ segments ($S$), the centers of the $L$ areas counting the two consecutive $L$'s, $LL$, as a single area (\{$L$,$LL$\}), the centers of $LL$'s only ($LL$), the centers of single isolated $L$'s surrounded by $S$'s on both sides ($SLS$). The boundaries between $L$ and $S$ segments, denoted henceforth as $|S|$ with the $|$'s symbolizing the edges, are displayed in the top panel (Edges). The edges $|S|$ can be categorized into two groups, namely  the edges between $LL$ and $S$ ($|LL|$) and between isolated $L$ and $S$ ($S|L|S$), which are shown by solid and dashed lines, respectively. Right: Schematic representation of the Fourier component (the ``average'' \textit{periodic} modulation) originating from the characteristic feature depicted to the left. Left: The inflation and the deflation of the Fibonacci sequence.
\label{FibonacciSelfSimSAve}}
\end{figure*}

In this section, we make an attempt to interpret the origin of the Fourier components observed in the CO\@. An important property of the Fibonacci sequence to be considered in this respect is its self-similarity. As depicted in the left panel of Fig.\ \ref{FibonacciSelfSimSAve}, the operation of the inflation rule (dividing an $L$ segment into two segments with the ratio of $\tau$:1 to get new $L$ and $S$, and renaming an $S$ segment as $L$) transforms the Fibonacci sequence to the Fibonacci sequence itself but scaled down by the factor $1/\tau$, while the deflation rule (the inverse process of the inflation) yields the Fibonacci sequence scaled up by the factor $\tau$. The self-similarity implies the presence in a FLSL of multiple length scales scaled to each other successively by the factor $\tau$, where we refer by the term ``length scale'' to the typical distance between characteristic features that take place repeatedly. The ratio $f_{j+1}/f_j$=$\tau$ between the adjacent major components $f_j$ can thus be interpreted in terms of the self-similarity.

An obvious example of such repeatedly occurring features is simply the segment $S$. Recalling that the CO results from the averaging of the square $v_\mathrm{d}^2$  of the drift velocity over the position of the guiding center $x_0$, it is the spatially averaged distance between the features that is reflected to the frequency $B_g$ of the CO\@. As depicted in Fig.\ \ref{FibonacciSelfSimSAve}, the distance between the centers of the adjacent $S$ segments is either $2L+S$=$\tau^3 S$ or $L+S$=$\tau^2 S$, reflecting the fact that $L$ segments appear in either of the two consecutive ($LL$) or single isolated ($L$) form. Noting that  the former (the latter) takes place in one-to-one correspondence with the $L$ ($S$) of the Fibonacci sequence retreating one generation by the deflation process, their relative abundance approaches $\tau$:1 for a long enough sequence. Therefore, the average distance between the centers reads, making use of the relation $\tau^2$=$\tau +1$,
\begin{equation}
\frac{(\tau^3 S) \tau +(\tau^2 S)}{\tau+1}=(\tau + 2)S=\frac{S}{f_3},
\end{equation}
accounting for the component $f_3$ in the CO, which is one of the dominant components in many of the FLSL samples (see Fig.\ \ref{ampcns}(a)). Since $LL$ is composed of seamless EB-resist in $L$-type samples (seamless open gap in $S$-type samples), $S/f_3$ represents as well the average distance between adjacent slabs of the resist or between adjacent free spaces. The values of $\langle \Delta x_\mathrm{res} \rangle$=$S/f_3$ for the samples examined in the present study are tabulated in Table \ref{Sampletbl}. Other principal components, $f_j$, can roughly be viewed as deriving from the inflation or deflation of $f_3$. More specifically, $f_1$, $f_2$, $f_4$, and $f_5$ are attributable to the centers of both types of the segments, the $L$ segments, the $LL$'s, and the isolated $L$'s, respectively (see Fig.\ \ref{FibonacciSelfSimSAve}). The minor components $f_{m,n}$ not categorized in $f_j$ can be ascribed to the edges of the resist slabs, which basically bear the frequency twice of the corresponding major components.

\begin{table}
\caption{The $x$-coordinate of the $k$-th appearance of characteristic features defined in Fig.\ \ref{FibonacciSelfSimSAve}. \label{kthpos}}
\begin{ruledtabular}
\begin{tabular}{ccccccc}
 & location (in unit $S$) \\
\hline
$x^{\{L,S\}}_k$ & $k/f_1-(1/2\tau)\{1+ \mathrm{Frac}[(k+1)/\tau]+\mathrm{Frac}(k/\tau)\}$ \\
$x^L_k$ & $k/f_2-\tau/2-\mathrm{Frac}(k/\tau)$ \\
$x^S_k$ & $k/f_3-1/2-\tau\, \mathrm{Frac}(k/\tau)$ \\
$x^{LL}_k$ & $k/f_4-\tau^2\, \mathrm{Frac}(k/\tau)$ \\
$x^{SLS}_k$ & $k/f_5+\tau/2-\tau^3\, \mathrm{Frac}(k/\tau)$ \\
$x^{|S|}_k$ & $k^+/f_{0,2}-\tau\, \mathrm{Frac}(k^+/2\tau)-0^+$ \\
$x^{|LL|}_k$ & $k^+/f_{2,-2}-\tau^2\, \mathrm{Frac}(k^+/2\tau)+(-1)^k\tau$ \\
$x^{S|L|S}_k$ & $k^+/f_{-2,4}-\tau^3\, \mathrm{Frac}(k^+/2\tau)+0^+\tau$ \\
\end{tabular}
\end{ruledtabular}
\end{table}

\begin{table}
\caption{The spatially averaged location for of $x_k$'s given in Table \ref{kthpos} for infinitely long sequence.  \label{avekthpos}}
\begin{ruledtabular}
\begin{tabular}{ccccccc}
 & location (in unit $S$) \\
\hline
$\langle x^{\{L,S\}}_k \rangle$ & $k/f_1-1/\tau$ \\
$\langle x^L_k \rangle$ & $k/f_2-\tau^2/2$ \\
$\langle x^S_k \rangle$ & $k/f_3-\tau^2/2$ \\
$\langle x^{LL}_k \rangle$ & $k/f_4-\tau^2/2$ \\
$\langle x^{SLS}_k \rangle$ & $k/f_5-\tau^2/2$ \\
$\langle x^{|S|}_k \rangle$ & $(k+1/2)/2f_3-\tau^2/2$ \\
$\langle x^{|LL|}_k \rangle$ & $(k+1/2)/2f_4-\tau^2/2$ \\
$\langle x^{S|L|S}_k \rangle$ & $(k+1/2)/2f_5-\tau^2/2$ \\
\end{tabular}
\end{ruledtabular}
\end{table}

To gain insight into the phase of the modulation, we need to take a closer look at the position of the characteristic features. Since the $k$-th appearance ($k$=1,2,3,...) of the $L$ segment ($S$ segment) takes place at the $a_k$-th ($b_k$-th) place in the whole sequence, where $a_k$=$k+\mathrm{Int}(k/\tau)$ and $b_k$=$2k+\mathrm{Int}(k/\tau)$ represent a pair of Beatty sequences \cite{Shroeder91} with $\mathrm{Int}(x)$ the integer part of $x$, the position of the $k$-th center of the $S$ segment, indicated by (red) vertical lines in the central panel of Fig.\ \ref{FibonacciSelfSimSAve}, is given by
\begin{equation}
x^S_k=(b_k-k) \tau S+ k S-\frac{S}{2}=\left[ \frac{k}{f_3}-\tau \mathrm{Frac}\left( \frac{k}{\tau} \right)-\frac{1}{2} \right] S,
\end{equation}
with $\mathrm{Frac}(x)$=$x-\mathrm{Int}(x)$ denoting the fractional part of $x$. The positions of other features can be written down along the same line, exploiting the correspondence of the feature to $L$ or $S$ in the inflated or deflated Fibonacci sequence, and tabulated in Table \ref{kthpos}, employing $S$ as the unit of the length. We used the notation $k^+ = k + [1-(-1)^k]/2$ in the description of the minor components, which takes the same number for an odd $k$ and the subsequent even $k$.

Since $\tau$ is an irrational number, $\mathrm{Frac}(k/\tau)$ and $\mathrm{Frac}(k^+/2\tau)$ take the value uniformly distributed in the interval $[0,1)$, fluctuating with $k$ around the central value $1/2$. The distance between the features under consideration, seen by the electrons traveling in the cyclotron orbit, fluctuates accordingly with the guiding center position $x_0$. In the context of the CO, we only need the average of the value over $x_0$. Therefore the fractional parts in Table \ref{kthpos} can be replaced by their average $1/2$, resulting in the spatially averaged locations $\langle x \rangle$ tabulated in Table \ref{avekthpos}. For the components deriving from the edges, we noted that $(-1)^k$ vanishes on averaging and hence $k^+$$\rightarrow$$k+1/2$, and also used the relations $f_{0,2}$=2$f_3$, $f_{2,-2}$=2$f_4$, and $f_{-2,4}$=2$f_5$.

\begin{figure}[tb]
\includegraphics[bbllx=40,bblly=40,bburx=520,bbury=810,width=8.5cm]{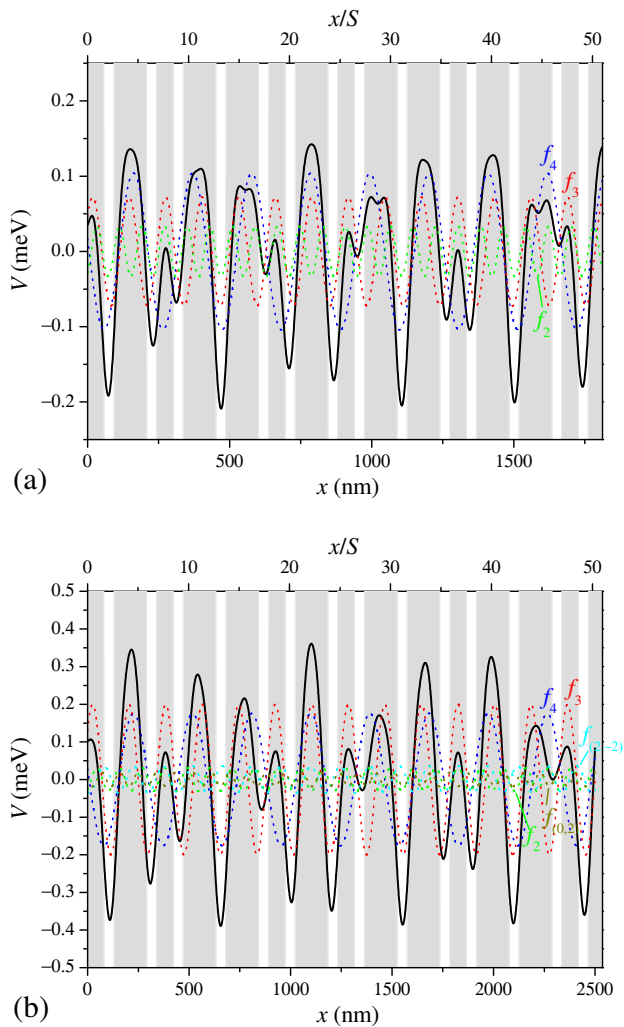}%
\caption{Potential profile of FLSLs (solid lines) reconstructed by the superposition of the Fourier components (dotted lines) with the amplitudes determined by the analysis of the CO and the phase obtained from the consideration of the origin of the components, for $L$-type samples 50$L$ (a) and 70$L$ (b). Note the difference in both vertical and horizontal scales. The horizontal axis is scaled so as to keep $x/S$ (top axis) the same for both figures. The shades indicate the area in the 2DEG plane located right beneath the resist, namely the $L$-sites. \label{profL}}
\end{figure}

\begin{figure}[tb]
\includegraphics[bbllx=40,bblly=40,bburx=520,bbury=810,width=8.5cm]{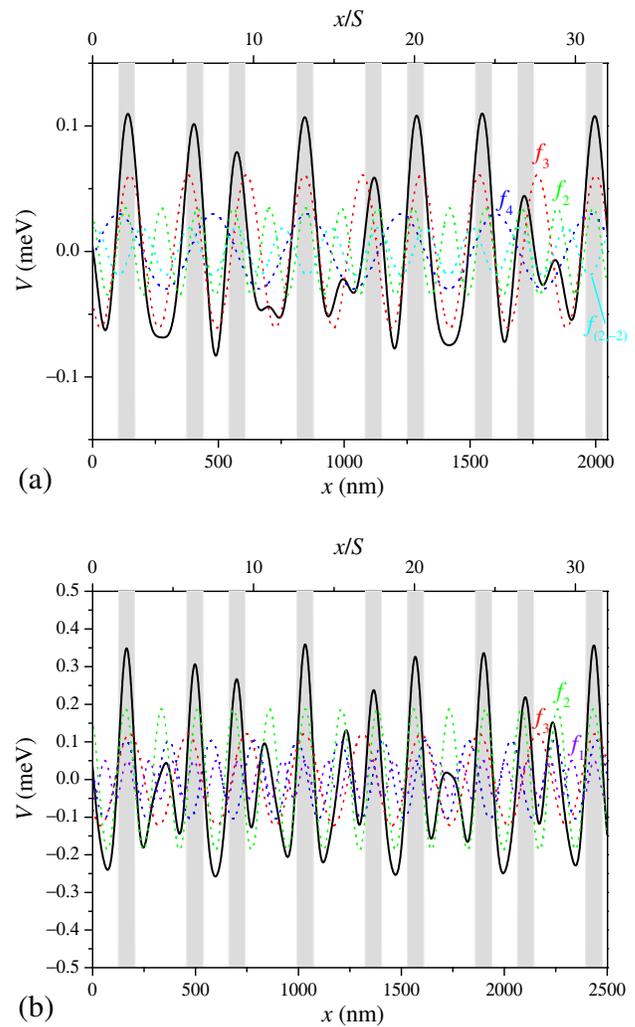}%
\caption{Similar to Fig.\ \ref{profL} for $S$-type samples 90$S$ (a) and 110$S$ (b). The shades indicate the $S$-sites. \label{profS}}
\end{figure}

For the reconstruction of the potential profile, it is necessary to further relate the characteristic features discussed thus far to the potential modulation they create. This requires detailed knowledge of the strain introduced into the GaAs/AlGaAs wafer by the resist slabs, which is beyond the reach of the present study. Here we simply assume that the potential energy becomes higher at the 2DEG areas directly beneath the resist slabs, namely at $L$-sites ($S$-sites) for $L$-type ($S$-type) samples, which are probably under compressive strain. Accordingly, the Fourier components including their phases are written, referring to Table \ref{avekthpos}, as $\pm V_j \cos [2\pi f_j (x/S+\tau^2/2)]$ for $j$=2, 4, 5, and $\mp V_3 \cos [2\pi f_3 (x/S+\tau^2/2)]$ for $j$=3, with the upper (lower) sign for $L$-type ($S$-type) samples. For the component $f_1$ which includes $L$-sites and $S$-sites alike, we presume that the $L$-sites dominate the sign: $\pm V_1 \cos [2\pi f_1 (x/S+1/\tau)]$. For edges, we assign the sign opposite to the area in between, namely, $\pm V_{0,2} \cos [4\pi f_3 (x/S+\tau/4)]$ and $\mp V_{2,-2} \cos [4\pi f_4 (x/S-1/4\tau)]$ (the component $f_{-2,4}$ was not observed in the present experiment).

The modulation profile is obtained by adding up all the components observed in the experiment. Examples of the profiles thus reconstructed, as well as the constituent Fourier components, are displayed in Figs.\ \ref{profL} and \ref{profS} for $L$-type (50$L$ and 70$L$) and $S$-type (90$S$ and 110$S$) samples, respectively, using the phases assumed above. It is intriguing to note that the superposition of a small number of incommensurate periodic sinusoidal modulations basically reproduce the quasiperiodic profile. Because of the low-pass filtering action of the attenuation $\exp(-gd)$=$\exp[-(2\pi f/S)d]$ mentioned earlier, higher frequency components (components $f_j$ with smaller $j$), i.e., the components with shorter length scale, are damped heavier for samples with smaller $S$. This is reflected in the difference of the profile between samples, resulting in rather dull line shape for samples with smaller $S$ when plotted against $x$ scaled by the unit length $S$ (see the top axes). The profile of $S$-type samples is qualitatively similar to that of $L$-type samples but laid up-side-down. Quantitatively, however, the modulation amplitude of $S$-type samples is much smaller than that of corresponding $L$-type samples with the same set of length units $L$ and $S$ (compare samples 90$L$ and 90$S$ in Fig.\ \ref{ampcns}). This is attributable to $1/\tau$ times smaller coverage by the resist slabs of the surface that exerts much smaller strain to the underlying GaAs/AlGaAs 2DEG wafer.
 
\section{Discussion \label{Discussion}}
What we have shown in the previous sections demonstrates that the Fourier analysis of the CO serves as a useful tool to probe the complicated profile of the modulated potential introduced into a 2DEG. FLSLs contain a number of Fourier components and therefore provide archetypal proving ground for the technique. The method, however, can be applied in wider range of systems that are comprised of multiple Fourier contents, e.g., periodic systems that retain higher harmonics. The technique is, in a sense, analogous to diffraction techniques that determine the crystal structures. In fact, Eq.\ (\ref{rhoadd}) with Eq.\ (\ref{magres1}) closely resembles the Patterson function \cite{Cowley81} $P(x)$=$\Sigma_g (|V_g|^2/2) \cos (gx)$ by replacing $x$=$2R_\mathrm{c}^*$=$2R_\mathrm{c}-\pi/2g$, and therefore can roughly be viewed as the autocorrelation function of the modulated potential measured by the cyclotron diameter with offset, $2 R_\mathrm{c}^*$. In this section, we discuss the limitations of the CO as a technique to determine the ``structure'' of the modulated potential.

Firstly, the information on the phase of the Fourier components is lost from the CO, as has been discussed so far. Again this resembles the situation in the diffraction techniques; the phases are unavailable in the diffraction pattern, which records only the intensity of the diffraction spots. In both cases, the phases have to be estimated from the symmetry of the system or from the physical origin of the component, as has been done in the preceding section. This can be a major source of uncertainty in the obtained reconstructed structures. Note, however, that the knowledge of the intensity of Fourier contents (without their phase interrelation) suffices for their explanation in many phenomena, a representative example being, of course, the CO itself.

Secondly, because of the thermal damping factor $A(T/T_g)$ in Eq.\ (\ref{magres1}) that decreases with increasing $g$, the CO is less sensitive to components with larger $g$, i.e., to shorter length scales. Therefore higher frequency components can be overlooked in CO traces, especially in those taken at higher temperatures. In principle, complementary use of the geometric resonance of the open orbit, \cite{Endo05N} which is a sensitive probe to higher frequency components, \cite{Endo06E} can fill the gap. However, the amplitude of the open orbit resonance still remains unexplained quantitatively, and its application to the determination of the modulation amplitude is yet to be explored. Conversely, the requirement for the electrons to travel at least one cycle of the cyclotron orbit before being scattered sets an ultimate upper limit, roughly equal to the mean free path, in the length scale that can be examined by the CO\@. Practical upper limit will be smaller than this, and will be reached when the oscillation resulting from the length scale becomes indistinguishable from the slowly-varying background (when the peak becomes indiscernible in the Fourier spectrum as shown in Fig.\ \ref{FB28a4FFT} due to the low-pass-filtering action of the operation of $d^2/dB^2$). Therefore, the modulation profile may possibly contain higher or lower frequency components that elude the detection by the CO\@.

Before concluding the paper, we briefly discuss the prospective use of a FLSL as a system to experimentally test the exotic properties in quasiperiodic systems that are theoretically predicted. \cite{Aubry80,Sokoloff85,Kohmoto87} The capability to experimentally determine the profile of the potential experienced by the electrons is obviously of great advantage to this end. However, further improvement of the sample that enables to control the electron density is desirable, since it will be necessary for many purposes to vary the Fermi energy or the Fermi wave number in detail. In 2DEG-based devices, this can generally be achieved by implementing a uniform gate. Note that in LSLs we have to resort to a back gate, since the front surface of the wafer is occupied by the grating that introduces the modulation. For back-gated LSLs, it was found that the modulation amplitude shows strong increase accompanying the decrease in the electron density by the application of the negative gate bias, \cite{Endo05MA} which might complicate the interpretation of the data obtained by the experiments. A project to develop LSLs with conducting resist gratings that can be biased to compensate for the change in the modulation amplitude induced by the back-gate bias is now under progress. Another caveat to be born in mind is that our FLSLs are basically a two-dimensional system where electrons can travel both parallel and perpendicular to the one-dimensional modulation. In many theories, by contrast, strictly one-dimensional systems are considered. It may therefore be necessary for some purposes to start the device fabrication from a quantum wire instead of a plain 2DEG\@.

\section{Conclusions \label{Conclusions}}
We have shown that the complicated CO observed in a FLSL can be decomposed into the constituent Fourier components by applying a numerical band-pass filter, with the aid of numerical operation of $d^2/dB^2$ for background elimination (and later integration by $B$ twice for recovery). From the amplitude of the component thus isolated, the magnitude $V_g$ of the potential modulation responsible for the component can be determined. The analysis reveals that the potential profile seen by the electrons in a FLSL is basically described by the superposition of a small number of incommensurate periodic modulations. The periods of the major components are given by $S/f_j$ with $f_j$ in Eq.\ (\ref{fj}), and are explicable in terms of inflation and deflation rules that lead to the self-similarity. By examining the physical origin of the components, we can estimate the phase of each component, which allows us to reconstruct the modulation profile. With the detailed knowledge of the potential profile, we believe that a FLSL provides an excellent arena to experimentally explore exotic phenomena (localization, critical  wave functions, a Cantor set spectrum, etc.) envisaged in the quasiperiodic systems.


%



\begin{acknowledgments}
This work was supported by Grant-in-Aid for Scientific Research (C) (18540312) and (A) (18204029) from the Ministry of Education, Culture, Sports, Science and Technology.
\end{acknowledgments}

\bibliography{lsls,qperiodic,NoteFiboCO,ourpps,misc}

\begin{thebibliography}{59}
\expandafter\ifx\csname natexlab\endcsname\relax\def\natexlab#1{#1}\fi
\expandafter\ifx\csname bibnamefont\endcsname\relax
  \def\bibnamefont#1{#1}\fi
\expandafter\ifx\csname bibfnamefont\endcsname\relax
  \def\bibfnamefont#1{#1}\fi
\expandafter\ifx\csname citenamefont\endcsname\relax
  \def\citenamefont#1{#1}\fi
\expandafter\ifx\csname url\endcsname\relax
  \def\url#1{\texttt{#1}}\fi
\expandafter\ifx\csname urlprefix\endcsname\relax\def\urlprefix{URL }\fi
\providecommand{\bibinfo}[2]{#2}
\providecommand{\eprint}[2][]{\url{#2}}

\bibitem[{\citenamefont{Beenakker and van Houten}(1991)}]{BeenakkerR91}
\bibinfo{author}{\bibfnamefont{C.~W.~J.} \bibnamefont{Beenakker}}
  \bibnamefont{and} \bibinfo{author}{\bibfnamefont{H.}~\bibnamefont{van
  Houten}}, in \emph{\bibinfo{booktitle}{Solid State Physics}}, edited by
  \bibinfo{editor}{\bibfnamefont{H.}~\bibnamefont{Ehrenreich}}
  \bibnamefont{and} \bibinfo{editor}{\bibfnamefont{D.}~\bibnamefont{Turnbull}}
  (\bibinfo{publisher}{Academic Press}, \bibinfo{address}{San Diego},
  \bibinfo{year}{1991}), vol.~\bibinfo{volume}{44}, p.~\bibinfo{pages}{1}.

\bibitem[{\citenamefont{Beton et~al.}(1990)\citenamefont{Beton, Alves, Main,
  Eaves, Dellow, Henini, Hughes, Beaumont, and Wilkinson}}]{Beton90P}
\bibinfo{author}{\bibfnamefont{P.~H.} \bibnamefont{Beton}},
  \bibinfo{author}{\bibfnamefont{E.~S.} \bibnamefont{Alves}},
  \bibinfo{author}{\bibfnamefont{P.~C.} \bibnamefont{Main}},
  \bibinfo{author}{\bibfnamefont{L.}~\bibnamefont{Eaves}},
  \bibinfo{author}{\bibfnamefont{M.~W.} \bibnamefont{Dellow}},
  \bibinfo{author}{\bibfnamefont{M.}~\bibnamefont{Henini}},
  \bibinfo{author}{\bibfnamefont{O.~H.} \bibnamefont{Hughes}},
  \bibinfo{author}{\bibfnamefont{S.~P.} \bibnamefont{Beaumont}},
  \bibnamefont{and} \bibinfo{author}{\bibfnamefont{C.~D.~W.}
  \bibnamefont{Wilkinson}}, \bibinfo{journal}{Phys.\ Rev.\ B}
  \textbf{\bibinfo{volume}{42}}, \bibinfo{pages}{9229} (\bibinfo{year}{1990}).

\bibitem[{\citenamefont{Weiss et~al.}(1989)\citenamefont{Weiss, v.~Klitzing,
  Ploog, and Weimann}}]{Weiss89}
\bibinfo{author}{\bibfnamefont{D.}~\bibnamefont{Weiss}},
  \bibinfo{author}{\bibfnamefont{K.}~\bibnamefont{v.~Klitzing}},
  \bibinfo{author}{\bibfnamefont{K.}~\bibnamefont{Ploog}}, \bibnamefont{and}
  \bibinfo{author}{\bibfnamefont{G.}~\bibnamefont{Weimann}},
  \bibinfo{journal}{Europhys.\ Lett.} \textbf{\bibinfo{volume}{8}},
  \bibinfo{pages}{179} (\bibinfo{year}{1989}).

\bibitem[{\citenamefont{Winkler et~al.}(1989)\citenamefont{Winkler, Kotthaus,
  and Ploog}}]{Winkler89}
\bibinfo{author}{\bibfnamefont{R.~W.} \bibnamefont{Winkler}},
  \bibinfo{author}{\bibfnamefont{J.~P.} \bibnamefont{Kotthaus}},
  \bibnamefont{and} \bibinfo{author}{\bibfnamefont{K.}~\bibnamefont{Ploog}},
  \bibinfo{journal}{Phys.\ Rev.\ Lett.} \textbf{\bibinfo{volume}{62}},
  \bibinfo{pages}{1177} (\bibinfo{year}{1989}).

\bibitem[{\citenamefont{Geim et~al.}(1992)\citenamefont{Geim, Taboryski,
  Kristensen, Dubonos, and Lindelof}}]{Geim92}
\bibinfo{author}{\bibfnamefont{A.~K.} \bibnamefont{Geim}},
  \bibinfo{author}{\bibfnamefont{R.}~\bibnamefont{Taboryski}},
  \bibinfo{author}{\bibfnamefont{A.}~\bibnamefont{Kristensen}},
  \bibinfo{author}{\bibfnamefont{S.~V.} \bibnamefont{Dubonos}},
  \bibnamefont{and} \bibinfo{author}{\bibfnamefont{P.~E.}
  \bibnamefont{Lindelof}}, \bibinfo{journal}{Phys.\ Rev.\ B}
  \textbf{\bibinfo{volume}{46}}, \bibinfo{pages}{4324} (\bibinfo{year}{1992}).

\bibitem[{\citenamefont{M\protect{\"u}ller
  et~al.}(1995)\citenamefont{M\protect{\"u}ller, Weiss, von Klitzing, Streda,
  and Weimann}}]{Muller95}
\bibinfo{author}{\bibfnamefont{G.}~\bibnamefont{M\protect{\"u}ller}},
  \bibinfo{author}{\bibfnamefont{D.}~\bibnamefont{Weiss}},
  \bibinfo{author}{\bibfnamefont{K.}~\bibnamefont{von Klitzing}},
  \bibinfo{author}{\bibfnamefont{P.}~\bibnamefont{Streda}}, \bibnamefont{and}
  \bibinfo{author}{\bibfnamefont{G.}~\bibnamefont{Weimann}},
  \bibinfo{journal}{Phys.\ Rev.\ B} \textbf{\bibinfo{volume}{51}},
  \bibinfo{pages}{10236} (\bibinfo{year}{1995}).

\bibitem[{\citenamefont{Tornow et~al.}(1996)\citenamefont{Tornow, Weiss,
  Manolescu, Menne, v.~Klitzing, and Weimann}}]{Tornow96}
\bibinfo{author}{\bibfnamefont{M.}~\bibnamefont{Tornow}},
  \bibinfo{author}{\bibfnamefont{D.}~\bibnamefont{Weiss}},
  \bibinfo{author}{\bibfnamefont{A.}~\bibnamefont{Manolescu}},
  \bibinfo{author}{\bibfnamefont{R.}~\bibnamefont{Menne}},
  \bibinfo{author}{\bibfnamefont{K.}~\bibnamefont{v.~Klitzing}},
  \bibnamefont{and} \bibinfo{author}{\bibfnamefont{G.}~\bibnamefont{Weimann}},
  \bibinfo{journal}{Phys.\ Rev.\ B} \textbf{\bibinfo{volume}{54}},
  \bibinfo{pages}{16397} (\bibinfo{year}{1996}).

\bibitem[{\citenamefont{Milton et~al.}(2000)\citenamefont{Milton, Emeleus,
  Lister, Davies, and Long}}]{Milton00}
\bibinfo{author}{\bibfnamefont{B.}~\bibnamefont{Milton}},
  \bibinfo{author}{\bibfnamefont{C.~J.} \bibnamefont{Emeleus}},
  \bibinfo{author}{\bibfnamefont{K.}~\bibnamefont{Lister}},
  \bibinfo{author}{\bibfnamefont{J.~H.} \bibnamefont{Davies}},
  \bibnamefont{and} \bibinfo{author}{\bibfnamefont{A.~R.} \bibnamefont{Long}},
  \bibinfo{journal}{Physica E} \textbf{\bibinfo{volume}{6}},
  \bibinfo{pages}{555} (\bibinfo{year}{2000}).

\bibitem[{\citenamefont{Endo and Iye}(2002)}]{Endo02f}
\bibinfo{author}{\bibfnamefont{A.}~\bibnamefont{Endo}} \bibnamefont{and}
  \bibinfo{author}{\bibfnamefont{Y.}~\bibnamefont{Iye}},
  \bibinfo{journal}{Phys.\ Rev.\ B} \textbf{\bibinfo{volume}{66}},
  \bibinfo{pages}{075333} (\bibinfo{year}{2002}).

\bibitem[{\citenamefont{Endo and Iye}(2004)}]{Endo04EP}
\bibinfo{author}{\bibfnamefont{A.}~\bibnamefont{Endo}} \bibnamefont{and}
  \bibinfo{author}{\bibfnamefont{Y.}~\bibnamefont{Iye}},
  \bibinfo{journal}{Physica E} \textbf{\bibinfo{volume}{22}},
  \bibinfo{pages}{122} (\bibinfo{year}{2004}).

\bibitem[{\citenamefont{Smet et~al.}(1999)\citenamefont{Smet, Jobst, von
  Klitzing, Weiss, Wegscheider, and Umansky}}]{Smet99c}
\bibinfo{author}{\bibfnamefont{J.~H.} \bibnamefont{Smet}},
  \bibinfo{author}{\bibfnamefont{S.}~\bibnamefont{Jobst}},
  \bibinfo{author}{\bibfnamefont{K.}~\bibnamefont{von Klitzing}},
  \bibinfo{author}{\bibfnamefont{D.}~\bibnamefont{Weiss}},
  \bibinfo{author}{\bibfnamefont{W.}~\bibnamefont{Wegscheider}},
  \bibnamefont{and} \bibinfo{author}{\bibfnamefont{V.}~\bibnamefont{Umansky}},
  \bibinfo{journal}{Phys.\ Rev.\ Lett.} \textbf{\bibinfo{volume}{83}},
  \bibinfo{pages}{2620} (\bibinfo{year}{1999}).

\bibitem[{\citenamefont{Willett et~al.}(1999)\citenamefont{Willett, West, and
  Pfeiffer}}]{Willett99c}
\bibinfo{author}{\bibfnamefont{R.~L.} \bibnamefont{Willett}},
  \bibinfo{author}{\bibfnamefont{K.~W.} \bibnamefont{West}}, \bibnamefont{and}
  \bibinfo{author}{\bibfnamefont{L.~N.} \bibnamefont{Pfeiffer}},
  \bibinfo{journal}{Phys.\ Rev.\ Lett.} \textbf{\bibinfo{volume}{83}},
  \bibinfo{pages}{2624} (\bibinfo{year}{1999}).

\bibitem[{\citenamefont{Endo et~al.}(2001)\citenamefont{Endo, Kawamura,
  Katsumoto, and Iye}}]{Endo01c}
\bibinfo{author}{\bibfnamefont{A.}~\bibnamefont{Endo}},
  \bibinfo{author}{\bibfnamefont{M.}~\bibnamefont{Kawamura}},
  \bibinfo{author}{\bibfnamefont{S.}~\bibnamefont{Katsumoto}},
  \bibnamefont{and} \bibinfo{author}{\bibfnamefont{Y.}~\bibnamefont{Iye}},
  \bibinfo{journal}{Phys.\ Rev.\ B} \textbf{\bibinfo{volume}{63}},
  \bibinfo{pages}{113310} (\bibinfo{year}{2001}).

\bibitem[{\citenamefont{Deutschmann et~al.}(2001)\citenamefont{Deutschmann,
  Wegscheider, Rother, Bichler, Abstreiter, Albrecht, and
  Smet}}]{Deutschmann01}
\bibinfo{author}{\bibfnamefont{R.~A.} \bibnamefont{Deutschmann}},
  \bibinfo{author}{\bibfnamefont{W.}~\bibnamefont{Wegscheider}},
  \bibinfo{author}{\bibfnamefont{M.}~\bibnamefont{Rother}},
  \bibinfo{author}{\bibfnamefont{M.}~\bibnamefont{Bichler}},
  \bibinfo{author}{\bibfnamefont{G.}~\bibnamefont{Abstreiter}},
  \bibinfo{author}{\bibfnamefont{C.}~\bibnamefont{Albrecht}}, \bibnamefont{and}
  \bibinfo{author}{\bibfnamefont{J.~H.} \bibnamefont{Smet}},
  \bibinfo{journal}{Phys.\ Rev.\ Lett.} \textbf{\bibinfo{volume}{86}},
  \bibinfo{pages}{1857} (\bibinfo{year}{2001}).

\bibitem[{\citenamefont{Endo and Iye}(2005{\natexlab{a}})}]{Endo05N}
\bibinfo{author}{\bibfnamefont{A.}~\bibnamefont{Endo}} \bibnamefont{and}
  \bibinfo{author}{\bibfnamefont{Y.}~\bibnamefont{Iye}},
  \bibinfo{journal}{Phys.\ Rev.\ B} \textbf{\bibinfo{volume}{71}},
  \bibinfo{pages}{081303(R)} (\bibinfo{year}{2005}{\natexlab{a}}).

\bibitem[{\citenamefont{Gerhardts et~al.}(1991)\citenamefont{Gerhardts, Weiss,
  and Wulf}}]{Gerhardts91}
\bibinfo{author}{\bibfnamefont{R.~R.} \bibnamefont{Gerhardts}},
  \bibinfo{author}{\bibfnamefont{D.}~\bibnamefont{Weiss}}, \bibnamefont{and}
  \bibinfo{author}{\bibfnamefont{U.}~\bibnamefont{Wulf}},
  \bibinfo{journal}{Phys.\ Rev.\ B} \textbf{\bibinfo{volume}{43}},
  \bibinfo{pages}{5192} (\bibinfo{year}{1991}).

\bibitem[{\citenamefont{Schl\protect{\"o}sser
  et~al.}(1996)\citenamefont{Schl\protect{\"o}sser, Ensslin, Kotthaus, and
  Holland}}]{Schlosser96}
\bibinfo{author}{\bibfnamefont{T.}~\bibnamefont{Schl\protect{\"o}sser}},
  \bibinfo{author}{\bibfnamefont{K.}~\bibnamefont{Ensslin}},
  \bibinfo{author}{\bibfnamefont{J.~P.} \bibnamefont{Kotthaus}},
  \bibnamefont{and} \bibinfo{author}{\bibfnamefont{M.}~\bibnamefont{Holland}},
  \bibinfo{journal}{Europhys.\ Lett.} \textbf{\bibinfo{volume}{33}},
  \bibinfo{pages}{683} (\bibinfo{year}{1996}).

\bibitem[{\citenamefont{Albrecht et~al.}(1999)\citenamefont{Albrecht, Smet,
  Weiss, von Klitzing, Hennig, Langenbuch, Suhrke, R\protect{\"o}ssler,
  Umansky, and Schweizer}}]{Albrecht99}
\bibinfo{author}{\bibfnamefont{C.}~\bibnamefont{Albrecht}},
  \bibinfo{author}{\bibfnamefont{J.~H.} \bibnamefont{Smet}},
  \bibinfo{author}{\bibfnamefont{D.}~\bibnamefont{Weiss}},
  \bibinfo{author}{\bibfnamefont{K.}~\bibnamefont{von Klitzing}},
  \bibinfo{author}{\bibfnamefont{R.}~\bibnamefont{Hennig}},
  \bibinfo{author}{\bibfnamefont{M.}~\bibnamefont{Langenbuch}},
  \bibinfo{author}{\bibfnamefont{M.}~\bibnamefont{Suhrke}},
  \bibinfo{author}{\bibfnamefont{U.}~\bibnamefont{R\protect{\"o}ssler}},
  \bibinfo{author}{\bibfnamefont{V.}~\bibnamefont{Umansky}}, \bibnamefont{and}
  \bibinfo{author}{\bibfnamefont{H.}~\bibnamefont{Schweizer}},
  \bibinfo{journal}{Phys.\ Rev.\ Lett.} \textbf{\bibinfo{volume}{83}},
  \bibinfo{pages}{2234} (\bibinfo{year}{1999}).

\bibitem[{\citenamefont{Geisler et~al.}(2004)\citenamefont{Geisler, Smet,
  Umansky, von Klitzing, Naundorf, Ketzmerick, and Schweizer}}]{Geisler04}
\bibinfo{author}{\bibfnamefont{M.~C.} \bibnamefont{Geisler}},
  \bibinfo{author}{\bibfnamefont{J.~H.} \bibnamefont{Smet}},
  \bibinfo{author}{\bibfnamefont{V.}~\bibnamefont{Umansky}},
  \bibinfo{author}{\bibfnamefont{K.}~\bibnamefont{von Klitzing}},
  \bibinfo{author}{\bibfnamefont{B.}~\bibnamefont{Naundorf}},
  \bibinfo{author}{\bibfnamefont{R.}~\bibnamefont{Ketzmerick}},
  \bibnamefont{and}
  \bibinfo{author}{\bibfnamefont{H.}~\bibnamefont{Schweizer}},
  \bibinfo{journal}{Phys.\ Rev.\ Lett.} \textbf{\bibinfo{volume}{92}},
  \bibinfo{pages}{256801} (\bibinfo{year}{2004}).

\bibitem[{\citenamefont{Chowdhury et~al.}(2004)\citenamefont{Chowdhury, Long,
  Skuras, Davies, Lister, Pennelli, and Stanley}}]{Chowdhury04}
\bibinfo{author}{\bibfnamefont{S.}~\bibnamefont{Chowdhury}},
  \bibinfo{author}{\bibfnamefont{A.~R.} \bibnamefont{Long}},
  \bibinfo{author}{\bibfnamefont{E.}~\bibnamefont{Skuras}},
  \bibinfo{author}{\bibfnamefont{J.~H.} \bibnamefont{Davies}},
  \bibinfo{author}{\bibfnamefont{K.}~\bibnamefont{Lister}},
  \bibinfo{author}{\bibfnamefont{G.}~\bibnamefont{Pennelli}}, \bibnamefont{and}
  \bibinfo{author}{\bibfnamefont{C.~R.} \bibnamefont{Stanley}},
  \bibinfo{journal}{Phys.\ Rev.\ B} \textbf{\bibinfo{volume}{69}},
  \bibinfo{pages}{035330} (\bibinfo{year}{2004}).

\bibitem[{Not({\natexlab{a}})}]{NoteRev}
\bibinfo{note}{See., e. g., E. Maci\protect{\'a}, Rep.\ Prog.\ Phys.
  \textbf{69}, 397 (2006), for a recent review.}

\bibitem[{\citenamefont{Aubry and Andre}(1980)}]{Aubry80}
\bibinfo{author}{\bibfnamefont{S.}~\bibnamefont{Aubry}} \bibnamefont{and}
  \bibinfo{author}{\bibfnamefont{G.}~\bibnamefont{Andre}},
  \bibinfo{journal}{Ann.\ Isr.\ Phys.\ Soc.} \textbf{\bibinfo{volume}{3}},
  \bibinfo{pages}{133} (\bibinfo{year}{1980}).

\bibitem[{\citenamefont{Sokoloff}(1985)}]{Sokoloff85}
\bibinfo{author}{\bibfnamefont{J.~B.} \bibnamefont{Sokoloff}},
  \bibinfo{journal}{Phys.\ Rep.} \textbf{\bibinfo{volume}{126}},
  \bibinfo{pages}{189} (\bibinfo{year}{1985}).

\bibitem[{\citenamefont{Kohmoto et~al.}(1987)\citenamefont{Kohmoto, Sutherland,
  and Tang}}]{Kohmoto87}
\bibinfo{author}{\bibfnamefont{M.}~\bibnamefont{Kohmoto}},
  \bibinfo{author}{\bibfnamefont{B.}~\bibnamefont{Sutherland}},
  \bibnamefont{and} \bibinfo{author}{\bibfnamefont{C.}~\bibnamefont{Tang}},
  \bibinfo{journal}{Phys.\ Rev.\ B} \textbf{\bibinfo{volume}{35}},
  \bibinfo{pages}{1020} (\bibinfo{year}{1987}).

\bibitem[{\citenamefont{Merlin et~al.}(1985)\citenamefont{Merlin, Bajema,
  Clarke, Juang, and Bhattacharya}}]{Merlin85}
\bibinfo{author}{\bibfnamefont{R.}~\bibnamefont{Merlin}},
  \bibinfo{author}{\bibfnamefont{K.}~\bibnamefont{Bajema}},
  \bibinfo{author}{\bibfnamefont{R.}~\bibnamefont{Clarke}},
  \bibinfo{author}{\bibfnamefont{F.~Y.} \bibnamefont{Juang}}, \bibnamefont{and}
  \bibinfo{author}{\bibfnamefont{P.~K.} \bibnamefont{Bhattacharya}},
  \bibinfo{journal}{Phys.\ Rev.\ Lett.} \textbf{\bibinfo{volume}{55}},
  \bibinfo{pages}{1768} (\bibinfo{year}{1985}).

\bibitem[{\citenamefont{Todd et~al.}(1986)\citenamefont{Todd, Merlin, Clarke,
  Mohanty, and Axe}}]{Todd86}
\bibinfo{author}{\bibfnamefont{J.}~\bibnamefont{Todd}},
  \bibinfo{author}{\bibfnamefont{R.}~\bibnamefont{Merlin}},
  \bibinfo{author}{\bibfnamefont{R.}~\bibnamefont{Clarke}},
  \bibinfo{author}{\bibfnamefont{K.~M.} \bibnamefont{Mohanty}},
  \bibnamefont{and} \bibinfo{author}{\bibfnamefont{J.~D.} \bibnamefont{Axe}},
  \bibinfo{journal}{Phys.\ Rev.\ Lett.} \textbf{\bibinfo{volume}{57}},
  \bibinfo{pages}{1157} (\bibinfo{year}{1986}).

\bibitem[{\citenamefont{Yamaguchi et~al.}(1990)\citenamefont{Yamaguchi, Saiki,
  Tada, Ninomiya, Misawa, Kobayashi, Gonokami, and Yao}}]{Yamaguchi90}
\bibinfo{author}{\bibfnamefont{A.~A.} \bibnamefont{Yamaguchi}},
  \bibinfo{author}{\bibfnamefont{T.}~\bibnamefont{Saiki}},
  \bibinfo{author}{\bibfnamefont{T.}~\bibnamefont{Tada}},
  \bibinfo{author}{\bibfnamefont{T.}~\bibnamefont{Ninomiya}},
  \bibinfo{author}{\bibfnamefont{K.}~\bibnamefont{Misawa}},
  \bibinfo{author}{\bibfnamefont{T.}~\bibnamefont{Kobayashi}},
  \bibinfo{author}{\bibfnamefont{M.~K.} \bibnamefont{Gonokami}},
  \bibnamefont{and} \bibinfo{author}{\bibfnamefont{T.}~\bibnamefont{Yao}},
  \bibinfo{journal}{Solid State Commun.} \textbf{\bibinfo{volume}{75}},
  \bibinfo{pages}{955} (\bibinfo{year}{1990}).

\bibitem[{\citenamefont{Gellermann et~al.}(1994)\citenamefont{Gellermann,
  Kohmoto, Sutherland, and Taylor}}]{Gellermann94}
\bibinfo{author}{\bibfnamefont{W.}~\bibnamefont{Gellermann}},
  \bibinfo{author}{\bibfnamefont{M.}~\bibnamefont{Kohmoto}},
  \bibinfo{author}{\bibfnamefont{B.}~\bibnamefont{Sutherland}},
  \bibnamefont{and} \bibinfo{author}{\bibfnamefont{P.~C.}
  \bibnamefont{Taylor}}, \bibinfo{journal}{Phys.\ Rev.\ Lett.}
  \textbf{\bibinfo{volume}{72}}, \bibinfo{pages}{633} (\bibinfo{year}{1994}).

\bibitem[{\citenamefont{Munzar et~al.}(1994)\citenamefont{Munzar, Bo\protect{\u
  c}\protect{\'a}ek, Huml\protect{\'i}\protect{\u c}ek, and Ploog}}]{Munzar94}
\bibinfo{author}{\bibfnamefont{D.}~\bibnamefont{Munzar}},
  \bibinfo{author}{\bibfnamefont{L.}~\bibnamefont{Bo\protect{\u
  c}\protect{\'a}ek}},
  \bibinfo{author}{\bibfnamefont{J.}~\bibnamefont{Huml\protect{\'i}\protect{\u
  c}ek}}, \bibnamefont{and}
  \bibinfo{author}{\bibfnamefont{K.}~\bibnamefont{Ploog}}, \bibinfo{journal}{J.
  Phys.: Condens. Matter} \textbf{\bibinfo{volume}{6}}, \bibinfo{pages}{4107}
  (\bibinfo{year}{1994}).

\bibitem[{\citenamefont{Toet et~al.}(1991)\citenamefont{Toet, Potemski, Wang,
  Maan, Tapfer, and Ploog}}]{Toet91}
\bibinfo{author}{\bibfnamefont{D.}~\bibnamefont{Toet}},
  \bibinfo{author}{\bibfnamefont{M.}~\bibnamefont{Potemski}},
  \bibinfo{author}{\bibfnamefont{Y.~Y.} \bibnamefont{Wang}},
  \bibinfo{author}{\bibfnamefont{J.~C.} \bibnamefont{Maan}},
  \bibinfo{author}{\bibfnamefont{L.}~\bibnamefont{Tapfer}}, \bibnamefont{and}
  \bibinfo{author}{\bibfnamefont{K.}~\bibnamefont{Ploog}},
  \bibinfo{journal}{Phys.\ Rev.\ Lett.} \textbf{\bibinfo{volume}{66}},
  \bibinfo{pages}{2128} (\bibinfo{year}{1991}).

\bibitem[{\citenamefont{Mikul\protect{\'i}k
  et~al.}(1995)\citenamefont{Mikul\protect{\'i}k, Hol\protect{\'y},
  Kub\protect{\v e}na, and Ploog}}]{Mikulik95}
\bibinfo{author}{\bibfnamefont{P.}~\bibnamefont{Mikul\protect{\'i}k}},
  \bibinfo{author}{\bibfnamefont{V.}~\bibnamefont{Hol\protect{\'y}}},
  \bibinfo{author}{\bibfnamefont{J.}~\bibnamefont{Kub\protect{\v e}na}},
  \bibnamefont{and} \bibinfo{author}{\bibfnamefont{K.}~\bibnamefont{Ploog}},
  \bibinfo{journal}{Acta Cryst.} \textbf{\bibinfo{volume}{A51}},
  \bibinfo{pages}{825} (\bibinfo{year}{1995}).

\bibitem[{\citenamefont{Zhu et~al.}(1989)\citenamefont{Zhu, Ming, and
  Jiang}}]{Zhu89}
\bibinfo{author}{\bibfnamefont{Y.~Y.} \bibnamefont{Zhu}},
  \bibinfo{author}{\bibfnamefont{N.~B.} \bibnamefont{Ming}}, \bibnamefont{and}
  \bibinfo{author}{\bibfnamefont{W.~H.} \bibnamefont{Jiang}},
  \bibinfo{journal}{Phys.\ Rev.\ B} \textbf{\bibinfo{volume}{40}},
  \bibinfo{pages}{8536} (\bibinfo{year}{1989}).

\bibitem[{\citenamefont{Macon et~al.}(1991)\citenamefont{Macon, Desideri, and
  Sornette}}]{Macon91}
\bibinfo{author}{\bibfnamefont{L.}~\bibnamefont{Macon}},
  \bibinfo{author}{\bibfnamefont{J.~P.} \bibnamefont{Desideri}},
  \bibnamefont{and} \bibinfo{author}{\bibfnamefont{D.}~\bibnamefont{Sornette}},
  \bibinfo{journal}{Phys.\ Rev.\ B} \textbf{\bibinfo{volume}{44}},
  \bibinfo{pages}{6755} (\bibinfo{year}{1991}).

\bibitem[{\citenamefont{Kono et~al.}(1991)\citenamefont{Kono, Nakada, Narahara,
  and Ootuka}}]{Kono91}
\bibinfo{author}{\bibfnamefont{K.}~\bibnamefont{Kono}},
  \bibinfo{author}{\bibfnamefont{S.}~\bibnamefont{Nakada}},
  \bibinfo{author}{\bibfnamefont{Y.}~\bibnamefont{Narahara}}, \bibnamefont{and}
  \bibinfo{author}{\bibfnamefont{Y.}~\bibnamefont{Ootuka}},
  \bibinfo{journal}{J.\ Phys.\ Soc.\ Jpn.} \textbf{\bibinfo{volume}{60}},
  \bibinfo{pages}{368} (\bibinfo{year}{1991}).

\bibitem[{\citenamefont{Katsumoto et~al.}(1993)\citenamefont{Katsumoto, Sano,
  and Kobayashi}}]{Katsumoto93}
\bibinfo{author}{\bibfnamefont{S.}~\bibnamefont{Katsumoto}},
  \bibinfo{author}{\bibfnamefont{N.}~\bibnamefont{Sano}}, \bibnamefont{and}
  \bibinfo{author}{\bibfnamefont{S.}~\bibnamefont{Kobayashi}},
  \bibinfo{journal}{Solid State Commun.} \textbf{\bibinfo{volume}{85}},
  \bibinfo{pages}{223} (\bibinfo{year}{1993}).

\bibitem[{\citenamefont{Smith et~al.}(1996)\citenamefont{Smith, Chao, Niu, and
  Shih}}]{Smith96}
\bibinfo{author}{\bibfnamefont{A.~R.} \bibnamefont{Smith}},
  \bibinfo{author}{\bibfnamefont{K.}~\bibnamefont{Chao}},
  \bibinfo{author}{\bibfnamefont{Q.}~\bibnamefont{Niu}}, \bibnamefont{and}
  \bibinfo{author}{\bibfnamefont{C.}~\bibnamefont{Shih}},
  \bibinfo{journal}{Science} \textbf{\bibinfo{volume}{273}},
  \bibinfo{pages}{226} (\bibinfo{year}{1996}).

\bibitem[{\citenamefont{Ebert et~al.}(1999)\citenamefont{Ebert, Chao, Niu, and
  Shih}}]{Ebert99}
\bibinfo{author}{\bibfnamefont{P.}~\bibnamefont{Ebert}},
  \bibinfo{author}{\bibfnamefont{K.~J.} \bibnamefont{Chao}},
  \bibinfo{author}{\bibfnamefont{Q.}~\bibnamefont{Niu}}, \bibnamefont{and}
  \bibinfo{author}{\bibfnamefont{C.~K.} \bibnamefont{Shih}},
  \bibinfo{journal}{Phys.\ Rev.\ Lett.} \textbf{\bibinfo{volume}{83}},
  \bibinfo{pages}{3222} (\bibinfo{year}{1999}).

\bibitem[{\citenamefont{Moras et~al.}(2006)\citenamefont{Moras, Theis, Ferrari,
  Gardonio, Fujii, Horn, and Carbone}}]{Moras06}
\bibinfo{author}{\bibfnamefont{P.}~\bibnamefont{Moras}},
  \bibinfo{author}{\bibfnamefont{W.}~\bibnamefont{Theis}},
  \bibinfo{author}{\bibfnamefont{L.}~\bibnamefont{Ferrari}},
  \bibinfo{author}{\bibfnamefont{S.}~\bibnamefont{Gardonio}},
  \bibinfo{author}{\bibfnamefont{J.}~\bibnamefont{Fujii}},
  \bibinfo{author}{\bibfnamefont{K.}~\bibnamefont{Horn}}, \bibnamefont{and}
  \bibinfo{author}{\bibfnamefont{C.}~\bibnamefont{Carbone}},
  \bibinfo{journal}{Phys.\ Rev.\ Lett.} \textbf{\bibinfo{volume}{96}},
  \bibinfo{pages}{156401} (\bibinfo{year}{2006}).

\bibitem[{\citenamefont{Eom et~al.}(2006)\citenamefont{Eom, Jiang, Yu, Shi,
  Niu, Ebert, and Shih}}]{Eom06}
\bibinfo{author}{\bibfnamefont{D.}~\bibnamefont{Eom}},
  \bibinfo{author}{\bibfnamefont{C.~S.} \bibnamefont{Jiang}},
  \bibinfo{author}{\bibfnamefont{H.~B.} \bibnamefont{Yu}},
  \bibinfo{author}{\bibfnamefont{J.}~\bibnamefont{Shi}},
  \bibinfo{author}{\bibfnamefont{Q.}~\bibnamefont{Niu}},
  \bibinfo{author}{\bibfnamefont{P.}~\bibnamefont{Ebert}}, \bibnamefont{and}
  \bibinfo{author}{\bibfnamefont{C.~K.} \bibnamefont{Shih}},
  \bibinfo{journal}{Phys.\ Rev.\ Lett.} \textbf{\bibinfo{volume}{97}},
  \bibinfo{pages}{206102} (\bibinfo{year}{2006}).

\bibitem[{\citenamefont{Endo and Iye}(2007)}]{Endo07I}
\bibinfo{author}{\bibfnamefont{A.}~\bibnamefont{Endo}} \bibnamefont{and}
  \bibinfo{author}{\bibfnamefont{Y.}~\bibnamefont{Iye}}, \bibinfo{journal}{AIP
  Conf.\ Proc.} \textbf{\bibinfo{volume}{893}}, \bibinfo{pages}{575}
  (\bibinfo{year}{2007}).

\bibitem[{\citenamefont{Endo and Iye}(2008)}]{Endo08E}
\bibinfo{author}{\bibfnamefont{A.}~\bibnamefont{Endo}} \bibnamefont{and}
  \bibinfo{author}{\bibfnamefont{Y.}~\bibnamefont{Iye}},
  \bibinfo{journal}{Physica E} \textbf{\bibinfo{volume}{40}},
  \bibinfo{pages}{1145} (\bibinfo{year}{2008}).

\bibitem[{\citenamefont{Endo et~al.}(2000)\citenamefont{Endo, Katsumoto, and
  Iye}}]{Endo00E}
\bibinfo{author}{\bibfnamefont{A.}~\bibnamefont{Endo}},
  \bibinfo{author}{\bibfnamefont{S.}~\bibnamefont{Katsumoto}},
  \bibnamefont{and} \bibinfo{author}{\bibfnamefont{Y.}~\bibnamefont{Iye}},
  \bibinfo{journal}{Phys.\ Rev.\ B} \textbf{\bibinfo{volume}{62}},
  \bibinfo{pages}{16761} (\bibinfo{year}{2000}).

\bibitem[{Not({\natexlab{b}})}]{NoteBsign}
\bibinfo{note}{The formula is for $B$$>$0. For $B$$<$0, $B$ should be replaced
  with $-B$ in this and other formulae presented henceforth.}

\bibitem[{\citenamefont{Beenakker}(1989)}]{Beenakker89}
\bibinfo{author}{\bibfnamefont{C.~W.~J.} \bibnamefont{Beenakker}},
  \bibinfo{journal}{Phys.\ Rev.\ Lett.} \textbf{\bibinfo{volume}{62}},
  \bibinfo{pages}{2020} (\bibinfo{year}{1989}).

\bibitem[{\citenamefont{Mirlin and W\protect{\"o}lfle}(1998)}]{Mirlin98}
\bibinfo{author}{\bibfnamefont{A.~D.} \bibnamefont{Mirlin}} \bibnamefont{and}
  \bibinfo{author}{\bibfnamefont{P.}~\bibnamefont{W\protect{\"o}lfle}},
  \bibinfo{journal}{Phys.\ Rev.\ B} \textbf{\bibinfo{volume}{58}},
  \bibinfo{pages}{12986} (\bibinfo{year}{1998}).

\bibitem[{\citenamefont{Zhang and Gerhardts}(1990)}]{Zhang90}
\bibinfo{author}{\bibfnamefont{C.}~\bibnamefont{Zhang}} \bibnamefont{and}
  \bibinfo{author}{\bibfnamefont{R.~R.} \bibnamefont{Gerhardts}},
  \bibinfo{journal}{Phys.\ Rev.\ B} \textbf{\bibinfo{volume}{41}},
  \bibinfo{pages}{12850} (\bibinfo{year}{1990}).

\bibitem[{\citenamefont{Peeters and Vasilopoulos}(1992)}]{Peeters92}
\bibinfo{author}{\bibfnamefont{F.~M.} \bibnamefont{Peeters}} \bibnamefont{and}
  \bibinfo{author}{\bibfnamefont{P.}~\bibnamefont{Vasilopoulos}},
  \bibinfo{journal}{Phys.\ Rev.\ B} \textbf{\bibinfo{volume}{46}},
  \bibinfo{pages}{4667} (\bibinfo{year}{1992}).

\bibitem[{\citenamefont{Gerhardts}(1992)}]{Gerhardts92}
\bibinfo{author}{\bibfnamefont{R.~R.} \bibnamefont{Gerhardts}},
  \bibinfo{journal}{Phys.\ Rev.\ B} \textbf{\bibinfo{volume}{45}},
  \bibinfo{pages}{3449} (\bibinfo{year}{1992}).

\bibitem[{\citenamefont{Endo and Iye}(2005{\natexlab{b}})}]{Endo05HH}
\bibinfo{author}{\bibfnamefont{A.}~\bibnamefont{Endo}} \bibnamefont{and}
  \bibinfo{author}{\bibfnamefont{Y.}~\bibnamefont{Iye}}, \bibinfo{journal}{J.\
  Phys.\ Soc.\ Jpn.} \textbf{\bibinfo{volume}{74}}, \bibinfo{pages}{2797}
  (\bibinfo{year}{2005}{\natexlab{b}}).

\bibitem[{\citenamefont{Skuras et~al.}(1997)\citenamefont{Skuras, Long, Larkin,
  Davies, and Holland}}]{Skuras97}
\bibinfo{author}{\bibfnamefont{E.}~\bibnamefont{Skuras}},
  \bibinfo{author}{\bibfnamefont{A.~R.} \bibnamefont{Long}},
  \bibinfo{author}{\bibfnamefont{I.~A.} \bibnamefont{Larkin}},
  \bibinfo{author}{\bibfnamefont{J.~H.} \bibnamefont{Davies}},
  \bibnamefont{and} \bibinfo{author}{\bibfnamefont{M.~C.}
  \bibnamefont{Holland}}, \bibinfo{journal}{Appl.\ Phys.\ Lett.}
  \textbf{\bibinfo{volume}{70}}, \bibinfo{pages}{871} (\bibinfo{year}{1997}).

\bibitem[{\citenamefont{Levine and Steinhardt}(1986)}]{Levine86}
\bibinfo{author}{\bibfnamefont{D.}~\bibnamefont{Levine}} \bibnamefont{and}
  \bibinfo{author}{\bibfnamefont{P.~J.} \bibnamefont{Steinhardt}},
  \bibinfo{journal}{Phys.\ Rev.\ B} \textbf{\bibinfo{volume}{34}},
  \bibinfo{pages}{596} (\bibinfo{year}{1986}).

\bibitem[{\citenamefont{Elser}(1986)}]{Elser86}
\bibinfo{author}{\bibfnamefont{V.}~\bibnamefont{Elser}}, \bibinfo{journal}{Acta
  Cryst.} \textbf{\bibinfo{volume}{A42}}, \bibinfo{pages}{36}
  (\bibinfo{year}{1986}).

\bibitem[{Not({\natexlab{c}})}]{NoteStr}
\bibinfo{note}{In a simple case where the lattice points in the original 2D
  square lattice are all identical, the structure factor reads \cite{Elser86}
  $S(z)$ $\propto$ $\sin(z)/z$ with $z$=$\pi \tau |-m+n\tau|/\sqrt{5}$ and
  therefore becomes large when $m/n$$\sim$$\tau$. Although the case does not
  exactly apply to our FLSLs, it hints at larger amplitude for $m$ and $n$
  given by two consecutive Fibonacci numbers.}

\bibitem[{\citenamefont{Dunlap}(1997)}]{Dunlap97}
\bibinfo{author}{\bibfnamefont{R.~A.} \bibnamefont{Dunlap}},
  \emph{\bibinfo{title}{The Golden Ratio and Fibonacci Numbers}}
  (\bibinfo{publisher}{World Scientific}, \bibinfo{address}{Singapore},
  \bibinfo{year}{1997}).

\bibitem[{\citenamefont{Posamentier and Lehmann}(2007)}]{Posamentier07}
\bibinfo{author}{\bibfnamefont{A.~S.} \bibnamefont{Posamentier}}
  \bibnamefont{and} \bibinfo{author}{\bibfnamefont{I.}~\bibnamefont{Lehmann}},
  \emph{\bibinfo{title}{The Fabulous Fibonacci Numbers}}
  (\bibinfo{publisher}{Prometheus Books}, \bibinfo{address}{New York},
  \bibinfo{year}{2007}).

\bibitem[{\citenamefont{Shroeder}(1991)}]{Shroeder91}
\bibinfo{author}{\bibfnamefont{M.~R.} \bibnamefont{Shroeder}},
  \emph{\bibinfo{title}{Fractals, Chaos, Power Laws}} (\bibinfo{publisher}{W.
  H. Freeman and Company}, \bibinfo{address}{New York}, \bibinfo{year}{1991}).

\bibitem[{\citenamefont{Cowley}(1981)}]{Cowley81}
\bibinfo{author}{\bibfnamefont{J.~M.} \bibnamefont{Cowley}},
  \emph{\bibinfo{title}{Diffraction Physics}}
  (\bibinfo{publisher}{North-Holland}, \bibinfo{address}{Amsterdam},
  \bibinfo{year}{1981}).

\bibitem[{\citenamefont{Endo and Iye}(2006)}]{Endo06E}
\bibinfo{author}{\bibfnamefont{A.}~\bibnamefont{Endo}} \bibnamefont{and}
  \bibinfo{author}{\bibfnamefont{Y.}~\bibnamefont{Iye}},
  \bibinfo{journal}{Physica E} \textbf{\bibinfo{volume}{34}},
  \bibinfo{pages}{640} (\bibinfo{year}{2006}).

\bibitem[{\citenamefont{Endo and Iye}(2005{\natexlab{c}})}]{Endo05MA}
\bibinfo{author}{\bibfnamefont{A.}~\bibnamefont{Endo}} \bibnamefont{and}
  \bibinfo{author}{\bibfnamefont{Y.}~\bibnamefont{Iye}}, \bibinfo{journal}{J.\
  Phys.\ Soc.\ Jpn.} \textbf{\bibinfo{volume}{74}}, \bibinfo{pages}{1792}
  (\bibinfo{year}{2005}{\natexlab{c}}).

\end{thebibliography}

\end{document}